\documentclass{vldb}
\usepackage{graphicx}
\usepackage{balance}  
\usepackage{times}
\usepackage{fancyvrb}
\usepackage{qtree}
\usepackage{tabularx}
\usepackage{url}
\usepackage{color}
\usepackage[all]{nowidow}

\newcommand{\CodeIn}[1]{{\small\texttt{#1}}}

\newcommand{\eat}[1] {}

\begin{document}

\newenvironment{observation}[1][Opportunity]{\begin{trivlist}
\item[\hskip \labelsep {\bfseries #1}]}{\end{trivlist}}


\title{Pregelix:  Big(ger) Graph Analytics on A Dataflow Engine} 

\vspace{-5ex}
\numberofauthors{1}
\author{
\alignauthor
Yingyi Bu$^\dagger$
\quad Vinayak Borkar$^\dagger$
\quad Jianfeng Jia$^\dagger$
\quad Michael J. Carey$^\dagger$
\quad Tyson Condie$^\ddagger$\\
       \affaddr{$^\dagger$University of California, Irvine\quad\quad $^\ddagger$University of California, Los Angeles}\\
       \email{$^\dagger${yingyib,vborkar,jianfenj,mjcarey}@ics.uci.edu, $^\ddagger$tcondie@cs.ucla.edu}
}

\maketitle

\vspace{-7ex}
\sloppy
\begin{abstract}

There is a growing need for distributed graph processing systems that are
capable of gracefully scaling to very large graph datasets.  Unfortunately,
this challenge has not been easily met due to the intense memory pressure
imposed by process-centric, message passing designs that many graph processing
systems follow.  Pregelix is a new open source distributed graph processing
system that is based on an iterative dataflow design that is better tuned to
handle both in-memory and out-of-core workloads.  As such, Pregelix offers
improved performance characteristics and scaling properties over current open
source systems (e.g., we have seen up to 15$\times$ speedup compared to Apache
Giraph and up to 35$\times$ speedup compared to distributed GraphLab), and
makes more effective use of available machine resources to support Big(ger)
Graph Analytics.

\end{abstract}

\section{Introduction }\label{introduction}

There are increasing demands to process Big Graphs for applications in social
networking (e.g., friend recommendations), the web (e.g., ranking pages), and
human genome assembly (e.g., extracting gene sequences).  Unfortunately, the
basic toolkits provided by first-generation ``Big Data'' Analytics platforms
(like Hadoop) lack an essential feature for Big Graph Analytics: MapReduce does not
support iteration (or equivalently, recursion) or certain key features required
to efficiently iterate ``around'' a MapReduce program.  Moreover, the MapReduce
programming model is not ideal for expressing many graph algorithms.  This
shortcoming has motivated several specialized approaches or libraries that
provide support for graph-based iterative programming on large clusters.

Google's Pregel is a prototypical example of such a platform; it allows
problem-solvers to ``think like a vertex'' by writing a few user-defined
functions (UDFs) that operate on vertices, which the framework can then apply
to an arbitrarily large graph in a parallel fashion.  Open source versions of
Pregel have since been developed in the systems community~\cite{Giraph, Hama}.
Perhaps unfortunately for both their implementors and users, each such platform
is a distinct new system that had to be built from the ground up.  Moreover,
these systems follow a process-centric design, in which a set of worker
processes are assigned partitions (containing sub-graphs) of the graph data and
scheduled across a machine cluster.  When a worker process launches, it reads
its assigned partition into memory, and executes the Pregel (message passing)
algorithm.  As we will see, such a design can suffer from poor support for
problems that are not memory resident.  Also desirable, would be the ability to
consider alternative runtime strategies that could offer more efficient
executions for different sorts of graph algorithms, datasets, and clusters.

The database community has spent nearly three decades building efficient
shared-nothing parallel query execution engines~\cite{ParallelDBMS} that
support out-of-core data processing operators (such as join and
group-by~\cite{QE}), and query optimizers~\cite{QO} that choose an ``optimal''
execution plan among different alternatives.  In addition, deductive database
systems---based on Datalog---were proposed to efficiently process recursive
queries~\cite{RecursiveQuery}, which can be used to solve graph problems such
as transitive closure.  However, there is no scalable implementation of Datalog
that offers the same fault-tolerant properties supported by today's ``Big
Data'' systems (e.g., \cite{Pregel, Hadoop, Nephele, Hyracks, Spark}).
Nevertheless, techniques for evaluating recursive queries---most notably
semi-na\"{i}ve evaluation---still apply and can be used to implement a
scalable, fault-tolerant Pregel runtime.

In this paper, we present Pregelix, a large-scale graph analytics system that
we began building in 2011.  Pregelix takes a novel set-oriented, iterative
dataflow approach to implementing the user-level Pregel programming model.  It
does so by treating the messages and vertex states in a Pregel computation like
tuples with a well-defined schema; it then uses database-style query evaluation
techniques to execute the user's program.
From a user's perspective, Pregelix provides the same Pregel programming
abstraction, just like Giraph~\cite{Giraph}.  However, from a runtime
perspective, Pregelix models Pregel's semantics as a logical query plan and
implements those semantics as an iterative dataflow of relational operators
that treat message exchange as a join followed by a group-by operation that
embeds functions that capture the user's Pregel program.  By taking this
approach, for the same logical plan, Pregelix is able to offer a set of
alternative physical evaluation strategies that can fit various workloads and
can be executed by Hyracks~\cite{Hyracks}, a general-purpose shared-nothing
dataflow engine (which is also the query execution engine for
AsterixDB~\cite{AsterixDB}).
By leveraging existing implementations of data-parallel operators and access
methods from Hyracks, we have avoided the need to build many critical system
components, e.g., bulk-data network transfer, out-of-core operator
implementations, buffer managers, index structures, and data shuffle.


To the best of our knowledge, Pregelix is the only Pregel-like system that
supports the full Pregel API, runs both in-memory workloads and out-of-core
workloads efficiently in a transparent manner on shared-nothing clusters, and
provides a rich set of runtime choices.  This paper makes the following
contributions:

\makeatletter
\def\topfigrule{\kern3\p@ \hrule \kern -3.4\p@} 
\def\botfigrule{\kern-3\p@ \hrule \kern 2.6\p@} 
\def\dblfigrule{\kern3\p@ \hrule \kern -3.4\p@} 
\makeatother \addtolength{\textfloatsep}{-.5\textfloatsep}
\addtolength{\dbltextfloatsep}{-.5\dbltextfloatsep}
\addtolength{\floatsep}{-.5\floatsep}
\addtolength{\dblfloatsep}{-.5\dblfloatsep}

\begin{list}{\labelitemi}{\leftmargin=1em}\itemsep 0pt \parskip 0pt
\item An analysis of existing Pregel-like systems: We revisit the Pregel programming
  abstraction and illustrate some shortcomings of typical custom-constructed
  Pregel-like systems (Section~\ref{background}).

\item A new Pregel architecture: We capture the semantics of Pregel in a
  logical query plan (Section~\ref{plan}), allowing us to execute Pregel as
  an iterative dataflow.

\item A system implementation: We first review the relevant building blocks in
  Hyracks (Section~\ref{hyracks}).  We then present the Pregelix system,
  elaborating the choices of data storage and physical plans as well as its key
  implementation details (Section~\ref{implementations}).

\item Case studies: We briefly describe several current use cases of Pregelix
  from our initial user community (Section~\ref{usecases}).

\item Experimental studies: We experimentally evaluate Pregelix in terms of
  execution time, scalability, throughput, plan
  flexibility, and implementation effort (Section~\ref{experiments}).

\end{list}

\section{Background and Problems}\label{background}
In this section, we first
briefly revisit the Pregel semantics and the Google Pregel runtime (Section~\ref{API}) as well as the
internals of Giraph, an open source Pregel-like system (Section~\ref{giraph}),
and then discuss the shortcomings of such custom-constructed Pregel architectures
(Section~\ref{problems}).

\subsection{Pregel Semantics and Runtime}\label{API}

Pregel~\cite{Pregel} was inspired by Valiant's bulk-synchronous parallel (BSP) model~\cite{BSP}. A Pregel program
describes a distributed graph algorithm in terms of vertices, edges, and a sequence of
iterations called supersteps.  The input to a Pregel computation is a directed
graph consisting of edges and vertices; each vertex is associated with a
mutable user-defined value and a boolean state indicating its liveness; each edge is associated with a source and
destination vertex and a mutable user-defined value.  During a superstep $S$, a
user-defined function (UDF) called \CodeIn{compute} is executed at each active vertex~$V$, and can perform any or all of the
following actions:
\begin{list}{\labelitemi}{\leftmargin=1em}\itemsep 0pt \parskip 0pt
  \item Read the messages sent to $V$ at the end of superstep~$S-1$;
  \item Generate messages for other vertices, which will be exchanged at the end of superstep~$S$;
  \item Modify the state of $V$ and its outgoing edges;
  \item Mutate the graph topology;
  \item Deactivate $V$ from the execution.
\end{list}
Initially, all vertices are in the active state.  A vertex can deactivate itself
by ``voting to halt'' in the call to \CodeIn{compute} using a Pregel provided
method.  A vertex is reactivated immediately if it receives a
message.  A Pregel program terminates when every vertex is in the inactive
state and no messages are in flight.


%
%

In a given superstep, any number of messages may be sent to a given
destination.  A user-defined \CodeIn{combine} function can be used to
pre-aggregate the messages for a destination. In addition, an aggregation function (e.g., min, max, sum, etc.) can be used
to compute a global aggregate among a set of participating vertices.  Finally,
the graph structure can be modified by any vertex; conflicts are handled by using
a partial ordering of operations such that all deletions go before insertions, and then by using a user-defined conflict resolution
function.

The Google Pregel runtime consists of a centralized master node that
coordinates superstep executions on a cluster of worker nodes.  
At the beginning of a Pregel job, each worker loads an assigned graph partition from a
distributed file system.  During execution, each worker 
calls the user-defined \CodeIn{compute} function on each active vertex in its
partition, passing in any messages sent to the vertex in the previous
superstep; outgoing messages are exchanged among workers.  The master is responsible for
coordinating supersteps and detecting termination.  Fault-tolerance is achieved
through checkpointing at user-specified superstep boundaries.

\subsection{Apache Giraph}\label{giraph}

\begin{figure}[!t]
  \centering
    \includegraphics[width=0.85\linewidth]{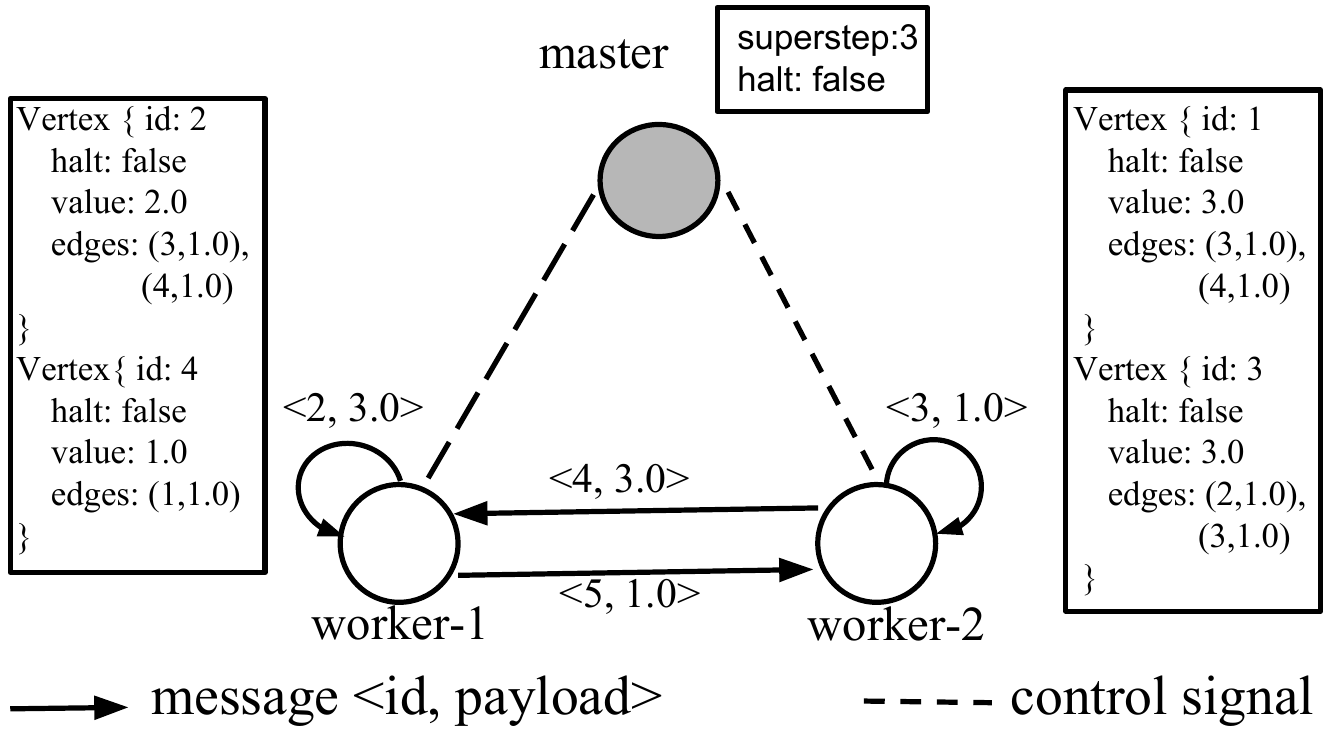}
    \vspace{-2ex}
    \caption{Giraph process-centric runtime.}\label{fig:giraph}
\end{figure}

Apache Giraph~\cite{Giraph} is an open source project that implements the
Pregel specification in Java on the Hadoop infrastructure.
Giraph launches master and worker instances in a Hadoop map-only
job\footnote{Alternatively, Giraph can use YARN~\cite{Yarn} for resource
allocation.}, where map tasks run master and worker instances.  Once started,
the master and worker map tasks internally execute the iterative computation
until completion, in a similar manner to Google's Pregel runtime.
Figure~\ref{fig:giraph} depicts Giraph's process-centric runtime for
implementing the Pregel programming model.
The vertex data is partitioned across worker tasks (two in this case).  Each
worker task communicates its control state (e.g., how many active vertices it
owns, when it has completed executing a given superstep, etc.) to the master
task.  The worker tasks establish communication channels between one another
for exchanging messages that get sent during individual vertex \CodeIn{compute}
calls; some of these messages could be for vertices on the same worker, e.g.,
messages $<$2, 3.0$>$ and $<$3,1.0$>$ in Figure~\ref{fig:giraph}.

\subsection{Issues and Opportunities}\label{problems}

Most process-centric Pregel-like systems have a minimum requirement for the
aggregate RAM needed to run a given algorithm on a particular dataset, making
them hard to configure for memory intensive computations and multi-user
workloads.  In fact, Google's Pregel only supports in-memory computations, as
stated in the original paper~\cite{Pregel}.  Hama~\cite{Hama} has limited
support for out-of-core vertex storage using immutable sorted files, but it
requires that the messages be memory-resident.  The latest version of Giraph
has preliminary out-of-core support; however, as we will see in
Section~\ref{experiments}, it does not yet work as expected.  Moreover, in the
Giraph user mailing
list\footnote{http://mail-archives.apache.org/mod\_mbox/giraph-user/} there are
26 cases (among 350 in total) of out-of-memory related issues from March 2013
to March 2014.  The users who posted those questions were typically from
academic institutes or small businesses that could not afford memory-rich
clusters, but who still wanted to analyze Big Graphs.  These issues essentially
stem from Giraph's ad-hoc, custom-constructed implementation of disk-based
graph processing.  This leads to our first opportunity to improve on the
current state-of-the-art.

\vspace{-1ex}
\begin{observation}
(Out-of-core Support)  Can we leverage mature database-style storage management and query evaluation techniques 
to provide better support for out-of-core workloads?
\end{observation}
\vspace{-1ex}


Another aspect of process-centric designs is that they only offer a single
physical layer implementation.  In those systems, the vertex storage strategy,
the message combination algorithm, the message redistribution strategy, and the
message delivery mechanism are each usually bound to one specific
implementation.  Therefore, we cannot choose between alternative implementation
strategies that would offer a better fit to a particular dataset, algorithm,
cluster or desktop.  For instance, the single source shortest paths algorithm exhibits
sparsity of messages, in which case a desired runtime strategy could avoid
iterating over all vertices by using an extra index to keep track of live
vertices.  This leads to our second opportunity.


\vspace{-1ex}
\begin{observation}
(Physical Flexibility) Can we better leverage data, algorithmic, and 
cluster/hardware properties to optimize a specific Pregel program?
\end{observation}
\vspace{-1ex}

The third issue is that the implementation of a process-centric runtime for the
Pregel model spans a full stack of network management, communication protocol,
vertex storage, message delivery and combination, memory management, and
fault-tolerance; the result is a complex (and hard-to-get-right) runtime system that
implements an elegantly simple Pregel semantics.  This leads to our third, software engineering
opportunity.

\vspace{-1ex}
\begin{observation}
(Software Simplicity) Can we leverage more from existing data-parallel 
platforms---platforms that have been improved for many years---to simplify the implementation
of a Pregel-like system?

\end{observation}
\vspace{-1ex}

We will see how these opportunities are exploited by our proposed architecture and implementation in Section~\ref{discussions}.

\section{The Pregel Logical Plan}\label{plan}
In this section, we model the semantics of Pregel as a logical query
plan. This model will guide the detailed design of the Pregelix system (Section~\ref{implementations}). 

\begin{table}[!t]
\begin{center}
\small
\begin{tabularx}{0.8\columnwidth}{|l||X|}
\hline Relation  &  Schema\\
\hline {\bf Vertex} & (vid, halt, value, edges)\\
\hline {\bf Msg} & (vid, payload)\\
\hline {\bf GS} & (halt, aggregate, superstep)\\
\hline
\end{tabularx}
\end{center}
\vspace{-4ex}
\caption{Nested relational schema that models the Pregel state.}
\label{table:relations}
\end{table}

\begin{figure}[!t]
  \centering
    \includegraphics[width=\linewidth]{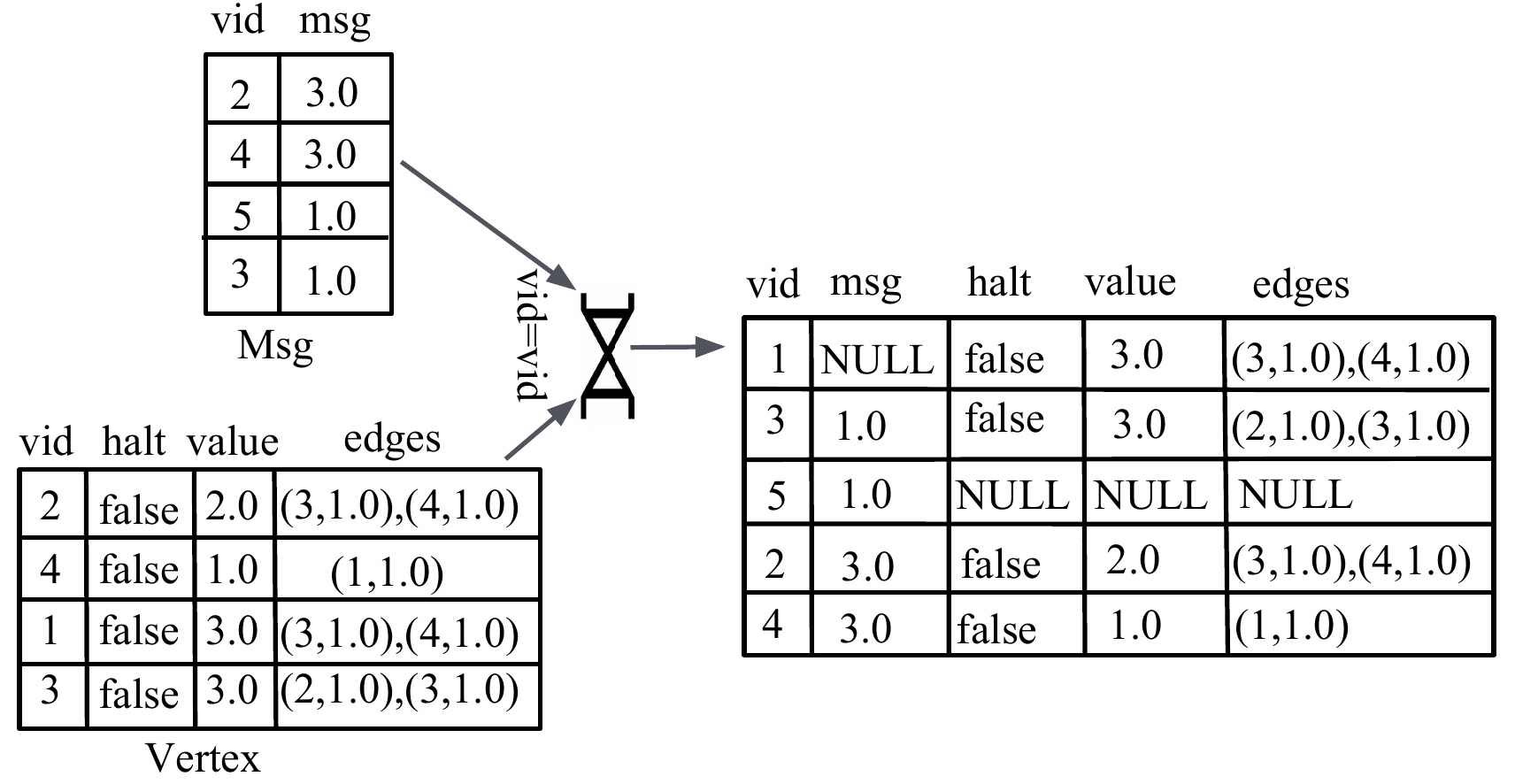}
    \vspace{-5ex}
    \caption{Implementing message-passing as a logical join.}\label{fig:joinmv}
\end{figure}

Our high level approach is to treat messages and vertices as data tuples
and use a join operation to model the message passing between vertices, as depicted in Figure~\ref{fig:joinmv}.
Table~\ref{table:relations} defines a set of nested relations that we use to model the
state of a Pregel execution.  The input data is modeled as an instance of
the \CodeIn{Vertex} relation; each row identifies a single vertex with its
halt, value, and edge states.  All vertices with a $halt=false$ state are
active in the current superstep.  The value and edges attributes represent the
vertex state and neighbor list, which can each be of a user-defined type.  The messages
exchanged between vertices in a superstep are modeled by an instance of the
\CodeIn{Msg} relation, which associates a destination vertex identifier with a
message payload.  Finally, the \CodeIn{GS} relation from Table~\ref{table:relations}
models the global state of the Pregel program; here, when $halt=true$ the program
terminates\footnote{This global halting state depends on the halting states of all vertices as well as the existence of messages.}, {\it aggregate} is a global state value, and {\it superstep} tracks the current
iteration count.

Figure~\ref{fig:joinmv} models message passing as a join between
the \CodeIn{Msg} and \CodeIn{Vertex} relations.  A full-outer-join is used to match messages with vertices corresponding to the Pregel semantics
as follows:
\begin{list}{\labelitemi}{\leftmargin=1em}\itemsep 0pt \parskip 0pt
  \item The inner case matches incoming messages with existing destination vertices;
  \item The left-outer case captures messages sent to vertices that may not exist; in this case,
     a vertex with the given {\it vid} is created with other fields set to NULL.
  \item The right-outer case captures vertices that have no messages; in this case, \CodeIn{compute} still needs 
    to be called for such a vertex if it is active. 
\end{list}
The output of the full-outer-join will be sent to further operator
processing steps that implement the Pregel semantics; some of these downstream
operators will involve UDFs that capture the details (e.g., \CodeIn{compute} implementation) of the given Pregel program.

\begin{table}[!t]
\begin{center}
\small
\begin{tabularx}{\columnwidth}{ |l||X|}
\hline UDF & Description \\
\hline {\bf compute} & Executed at each active vertex in every superstep. \\
\hline {\bf combine} & Aggregation function for messages.\\
\hline {\bf aggregate} & Aggregation function for the global state.\\
\hline {\bf resolve} & Used to resolve conflicts in graph mutations.\\
\hline
\end{tabularx}
\end{center}
\vspace{-4ex}
\caption{UDFs used to capture a Pregel program.}
\label{table:functions}
\end{table}

Table~\ref{table:functions} lists the UDFs that implement a given Pregel program.
In a given superstep, each active vertex is processed through a call to the
\CodeIn{compute} UDF, which is passed the messages sent to the vertex in the
previous superstep. The output of a call to \CodeIn{compute} is a tuple that
contains the following fields:
\begin{list}{\labelitemi}{\leftmargin=1em}\itemsep 0pt \parskip 0pt
  \item The possibly updated \CodeIn{Vertex} tuple. 
  \item A list of outbound messages (delivered in the next superstep).
  \item The global {\it halt} state contribution, which is {\it true} when the outbound message list is empty and the {\it halt} field of the updated vertex is {\it true}, and {\it false} otherwise.
  \item The global {\it aggregate} state contribution (tuples nested in bag).
  \item The graph mutations (a nested bag of tuples to insert/delete to/from the \CodeIn{Vertex} relation). 
\end{list}
As we will see below, this output is routed to downstream operators that
extract (project) one or more of these fields and execute the dataflow of a
superstep.  For instance, output messages are grouped by the destination vertex
id and aggregated by the \CodeIn{combine} UDF.  The global aggregate state contributions
of all vertices are passed to the \CodeIn{aggregate} function, which produces
the global aggregate state value for the subsequent superstep.  Finally, the \CodeIn{resolve} UDF accepts all
graph mutations---expressed as insertion/deletion tuples against the
\CodeIn{Vertex} relation---as input, and it resolves any conflicts before they are
applied to the \CodeIn{Vertex} relation.

\begin{figure}[!t]
  \centering
    \includegraphics[width=0.90\linewidth]{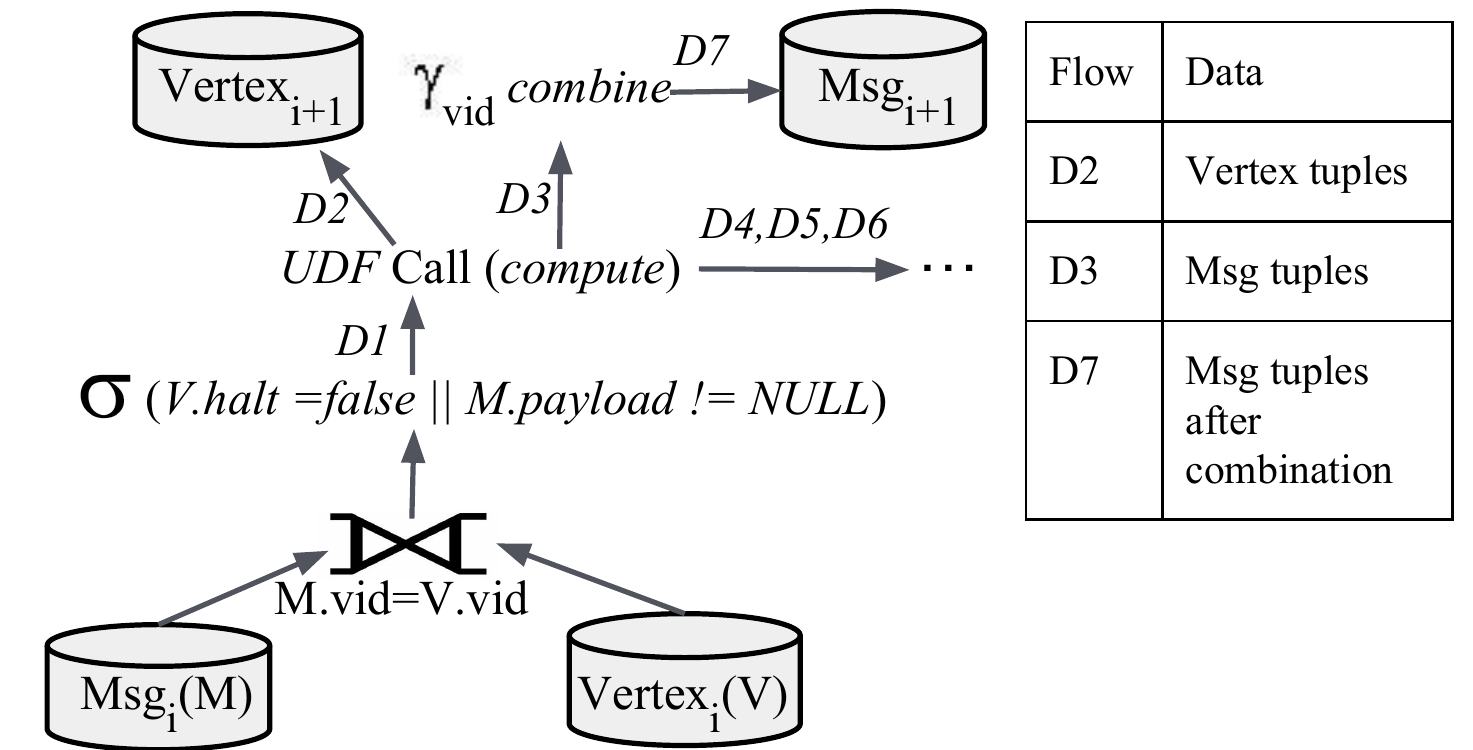}
    \vspace{-2ex}
    \caption{The basic logical query plan of a Pregel superstep $i$ which
reads the data generated from the last superstep (e.g., \CodeIn{Vertex$_{i}$}, \CodeIn{Msg$_{i}$}, and \CodeIn{GS$_{i}$}) and 
produces the data (e.g., \CodeIn{Vertex$_{i+1}$}, \CodeIn{Msg$_{i+1}$}, and \CodeIn{GS$_{i+1}$})  for superstep $i+1$. 
Global aggregation and synchronization are in Figure~\ref{fig:gs}, and vertex addition and removal are in Figure~\ref{fig:addremove}.}\label{fig:logicalplan}
\end{figure}

We now turn to the description of a single logical dataflow plan; we divide it
into three figures that each focus on a specific application of the (shared) output of the
\CodeIn{compute} function.  The relevant dataflows are labeled in each figure.
Figure~\ref{fig:logicalplan} defines the input to the \CodeIn{compute} UDF, the
output messages, and updated vertices.  Flow $D1$ describes the
\CodeIn{compute} input for superstep~$i$ as being the output of a
full-outer-join between \CodeIn{Msg} and \CodeIn{Vertex} (as described by
Figure~\ref{fig:joinmv}) followed by a selection predicate that prunes inactive
vertices.  The \CodeIn{compute} output pertaining to vertex data is projected
onto dataflow $D2$, which then updates the \CodeIn{Vertex} relation.  In
datafow $D3$, the message output is grouped by destination vertex id and
aggregated by the \CodeIn{combine} function\footnote{The default
\CodeIn{combine} gathers all messages for a given destination into a list.},
which produces flow $D7$ that is inserted into the \CodeIn{Msg} relation.

\begin{figure}[!t]
  \centering
    \includegraphics[width=\linewidth]{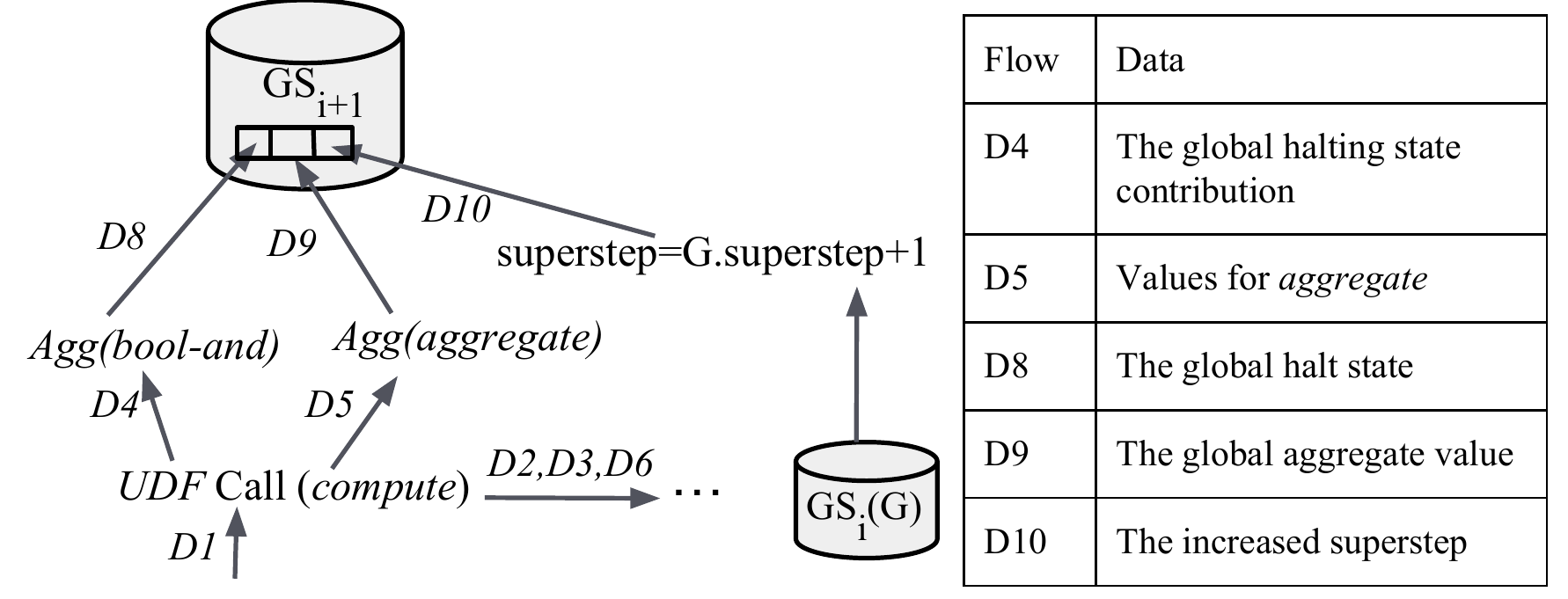}
    \vspace{-3ex}
    \caption{The plan segment that revises the global state.}\label{fig:gs}
\end{figure}

The global state relation \CodeIn{GS} contains a single tuple whose fields
comprise the global state.  Figure~\ref{fig:gs} describes the flows that
revise these fields in each superstep.  The halt state and global aggregate
fields depend on the output of \CodeIn{compute}, while the superstep
counter is simply its previous value plus one.  Flow $D4$ applies a
boolean aggregate function (logical AND) to the global halting state contribution from each vertex;
the output (flow $D8$) indicates the global halt state, which controls the
execution of another superstep.  Flow $D5$ routes the global aggregate state
contributions from all active vertices to the \CodeIn{aggregate} UDF which then
produces the global aggregate value (flow $D9$) for the next superstep.

\begin{figure}[!t]
  \centering
    \includegraphics[width=0.68\linewidth]{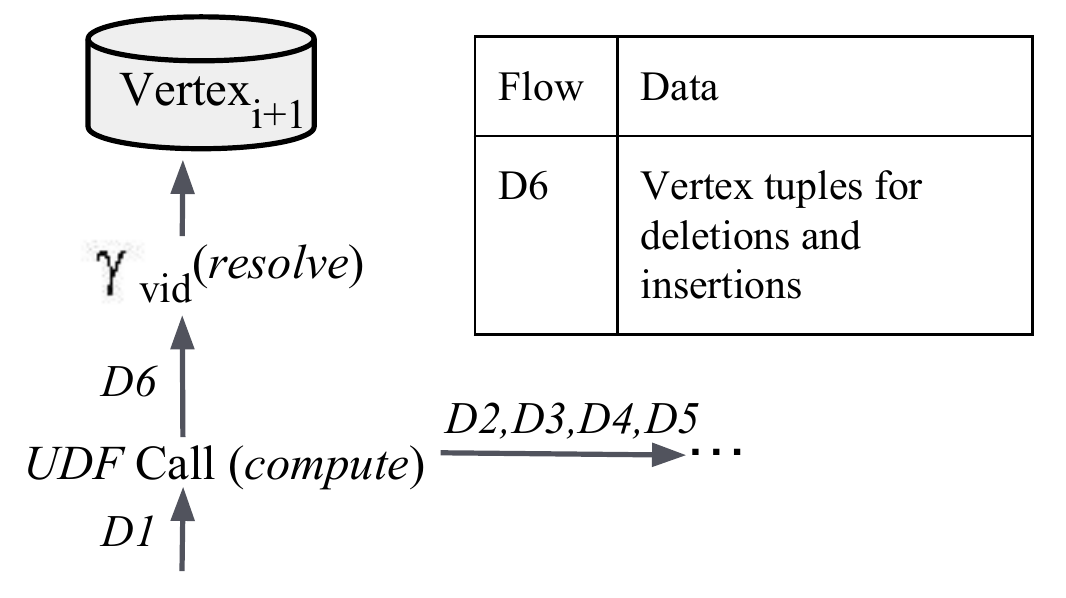}
    \vspace{-3ex}
    \caption{The plan segment for vertex addition/removal.}\label{fig:addremove}
\end{figure}

Graph mutations are specified by a \CodeIn{Vertex} tuple with an operation that
indicates insertion (adding a new vertex) or deletion (removing a
vertex)\footnote{Pregelix leaves the application-specific vertex deletion semantics
in terms of integrity constraints to application programmers.}.  Flow $D6$ in
Figure~\ref{fig:addremove} groups these mutation tuples by vertex id and
applies the \CodeIn{resolve} function to each group.  The output is then
applied to the \CodeIn{Vertex} relation.

\section{The Runtime Platform}\label{hyracks}
The Pregel logical plan could be implemented on any 
parallel dataflow engine, including Stratosphere~\cite{Nephele}, Spark~\cite{Spark}
or Hyracks~\cite{Hyracks}.
As we argue below, we believe that Hyracks is particularly well-suited for this style of computation; 
this belief is supported by Section~\ref{experiments}'s experimental results
(where some of the systems studied are based on other platforms).
The rest of this section covers the Hyracks platform~\cite{Hyracks}, which is Pregelix's
target runtime for the logical plan in Section~\ref{plan}.  Hyracks is a
data-parallel runtime in the same general space as Hadoop~\cite{Hadoop} and
Dryad~\cite{Dryad}.  Jobs are submitted to Hyracks in the form of DAGs
(directed acyclic graphs) that are made up of operators and connectors.
Operators are responsible for consuming input partitions and producing output
partitions.  Connectors perform redistributions of data between operators.  For
a submitted job, in a Hyracks cluster, a master machine dictates a set of
worker machines to execute clones of the operator DAG in parallel and
orchestrates data exchanges.
Below, we enumerate the features and components of Hyracks that we leverage to
implement the logical plan described in Section~\ref{plan}.

{\bf User-configurable task scheduling.}  The Hyracks engine allows users to express
task scheduling constraints (e.g., count constraints, location choice
constraints, or absolute location constraints) for each physical operator.  The
task scheduler of Hyracks is a constraint solver that comes up with a schedule
satisfying the user-defined constraints.  In Section~\ref{scheduling}, we
leverage this feature of Hyracks to implement sticky, iterative dataflows.

{\bf Access methods.}  B-trees and LSM B-trees are part of the Hyracks storage
library.  A B-tree~\cite{BTree} is a commonly used index structure in most
commercial databases; it supports efficient lookup and scan operations, but
a single tree update can cause random I/Os.  In contrast, the LSM
B-tree~\cite{LSM} puts updates into an in-memory component (e.g., an in-memory
B-tree); it merges the in-memory component with disk components in a
periodic manner, which turns random I/Os for updates into sequential ones.  The
LSM B-tree thus allows fast updates but may result in slightly slower lookups.

{\bf Group-by operators.}  The Hyracks operator library includes three group-by
operator implementations: sort-based group-by, which pushes group-by
aggregations into both the in-memory sort phase and the merge phase of an
external sort operator; HashSort group-by, which does the same thing as the
sort-based one except using hash-based group-by for the in-memory processing
phase; and preclustered group-by, which assumes incoming tuples are already
clustered by the group-by key and hence just applies the aggregation operation
in sequence to one group after the other.

{\bf Join operators.} Merge-based index full outer join and probe-based index
left-outer join are supported in the Hyracks operator library.  The full outer
join operator merges sorted input from the outer relation with an index scan on
the inner relation; tuples containing \CodeIn{NULL} values for missing fields will be
generated for no-matches.  The left-outer join operator,
for each input tuple in the outer relation, consults an index on the inner relation for
matches that produce join results or for no-matches that produce tuples with \CodeIn{NULL}
values for the inner relation attributes.

{\bf Connectors.} Hyracks connectors define inter-operator data exchange
patterns.  Here, we focus on the following three communication patterns: an
m-to-n partitioning connector repartitions the data based on a user-defined
partitioning function from m (sender-side) partitions to n (receiver-side)
partitions; the m-to-n partitioning merging connector does the same thing but
assumes tuples from the sender-side are ordered and therefore simply merges the
input streams at the receiver-side; the aggregator connector reduces all input
streams to a single receiver partition.

{\bf Materialization policies.} We use two materialization policies
that Hyracks supports for customizing connectors: fully pipelined, where
the data from a producer is immediately pushed to the consumer, and sender-side
materializing pipelined, where the data transfer channel launches two
threads at the sender side, one that writes output data to a local temporary file,
and another that pulls written data from the file and sends it to the receiver-side.

\section{The Pregelix System}\label{implementations}

In this section, we describe our implementation of the logical plan
(Section~\ref{plan}) on the Hyracks runtime (Section~\ref{hyracks}), which 
is core of the Pregelix system.  We elaborate on data-parallel execution
(Section~\ref{parallelism}), data storage (Section~\ref{storages}), physical
query plan alternatives (Section~\ref{physicalplans}), memory management
(Section~\ref{ooc}), fault-tolerance (Section~\ref{ft}), and job pipelining
(Section~\ref{pipelining}).  We conclude by summarizing the software components
of Pregelix (Section~\ref{overview}) and revisiting our three opportunities
(Section~\ref{discussions}).

\subsection{Parallelism}\label{parallelism}

\begin{figure}[!t]
  \centering
    \includegraphics[width=\linewidth]{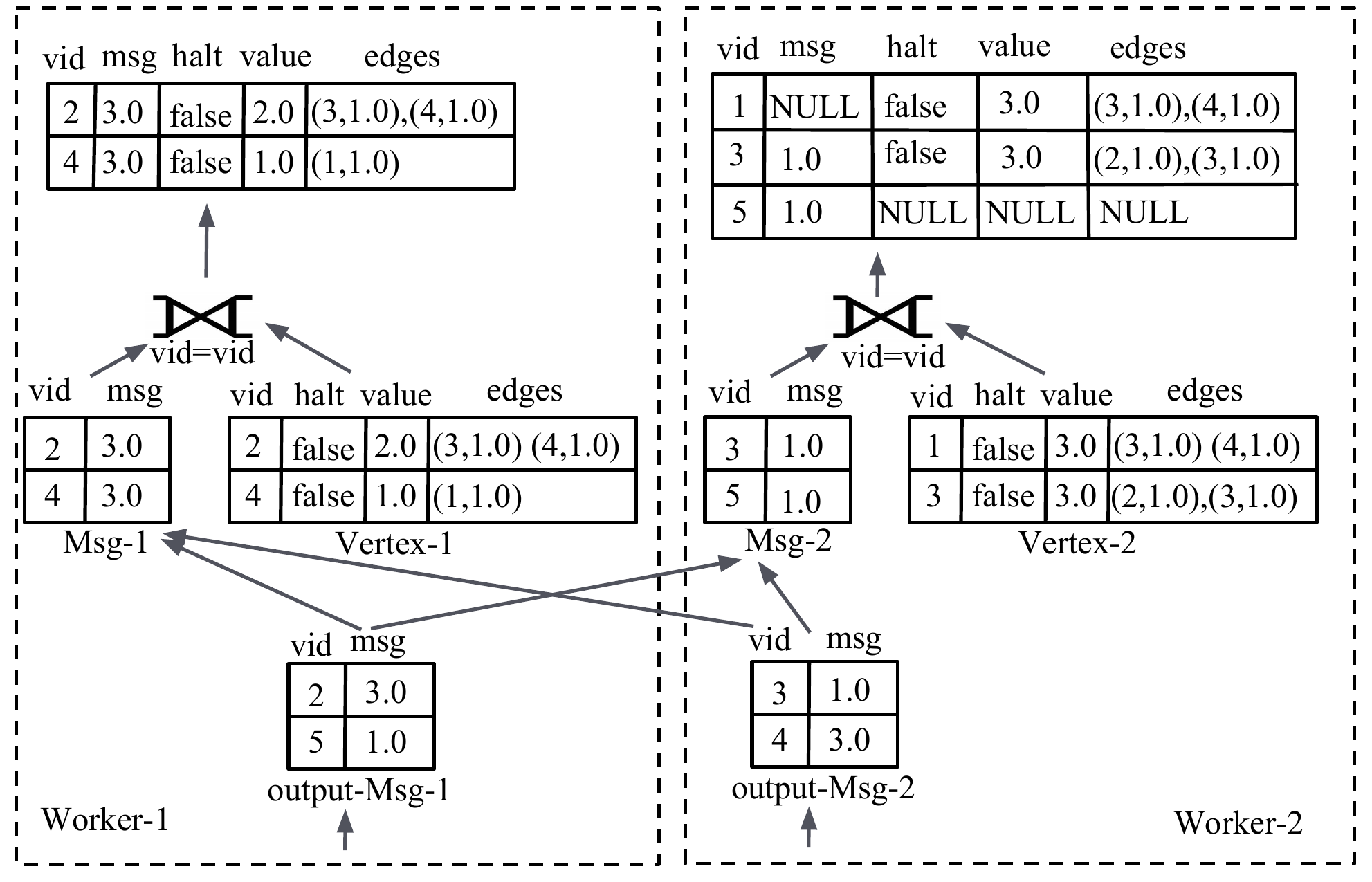}
    \vspace{-4ex}
    \caption{The parallelized join for the logical join in Figure~\ref{fig:joinmv}.}\label{fig:joinp}
\end{figure}

To parallelize the logical plan of the Pregel computation described in
Section~\ref{plan} at runtime, one or more clones of a physical plan---that
implements the logical plan---are shipped to Hyracks worker machines that run
in parallel.  Each clone deals with a single data partition.  During execution,
data is exchanged from the clones of an upstream operator to those of
a downstream operator through a Hyracks connector.  Figure~\ref{fig:joinp} shows
an example, where the logical join described in Figure~\ref{fig:joinmv} is
parallelized onto two workers and message tuples are exchanged from producer
partitions (operator clones) to consumer partitions (operator clones) using an
m-to-n partitioning connector, where m and n are equal to two.

\subsection{Data Storage}\label{storages}

Given a graph analytical job, Pregelix first loads the input graph dataset (the
initial \CodeIn{Vertex} relation) from a distributed file system, i.e., HDFS,
into a Hyracks cluster, partitioning it by {\it vid} using a user-defined
partitioning function\footnote{By default, we use hash partitioning.} across the
worker machines.  After the eventual completion of the overall Pregel computation,
the partitioned \CodeIn{Vertex} relation is scanned and dumped back to HDFS.
During the supersteps, at each worker node, one (or more) local indexes---keyed
off of the {\it vid} field---are used to store one (or more) partitions of the
\CodeIn{Vertex} relation.  Pregelix leverages both B-tree and LSM B-tree index
structures from the Hyracks storage library to store partitions of
\CodeIn{Vertex} on worker machines.  The choice of which index structure to use
is workload-dependent and user-selectable.  A B-tree index performs well on
jobs that frequently update vertex data in-place, e.g., PageRank.  An LSM
B-tree index performs well when the size of vertex data is changed drastically
from superstep to superstep, or when the algorithm performs frequent graph
mutations, e.g., the path merging algorithm in genome assemblers~\cite{Velvet}.

The \CodeIn{Msg} relation is initially empty; it is refreshed at the end of a
superstep with the result of the message \CodeIn{combine} function call in the
(logical) dataflow~$D7$ of Figure~\ref{fig:logicalplan}; the physical plan is
described in Section~\ref{groupbys}.  The message data is partitioned by
destination vertex id ({\it vid}) using the same partitioning function
applied to the vertex data, and is thus stored (in temporary local files)
on worker nodes that maintain the destination vertex data.  Furthermore, each
message partition is sorted by the {\it vid} field.

Lastly, we leverage HDFS to store the global state of a Pregelix job; an access
method is used to read and cache the global state at worker nodes when it is
referenced by user-defined functions like \CodeIn{compute}.

\subsection{Physical Query Plans}\label{physicalplans}

In this subsection, we dive into the details of the physical plans for the
logical plan described in Figures~\ref{fig:logicalplan}, \ref{fig:gs}, and
\ref{fig:addremove}.  Our discussion will cover message combination and
delivery, global states, graph mutations, and data redistribution.

\subsubsection{Message Combination}\label{groupbys}
\begin{figure}[!t]
  \centering
    \includegraphics[width=\linewidth]{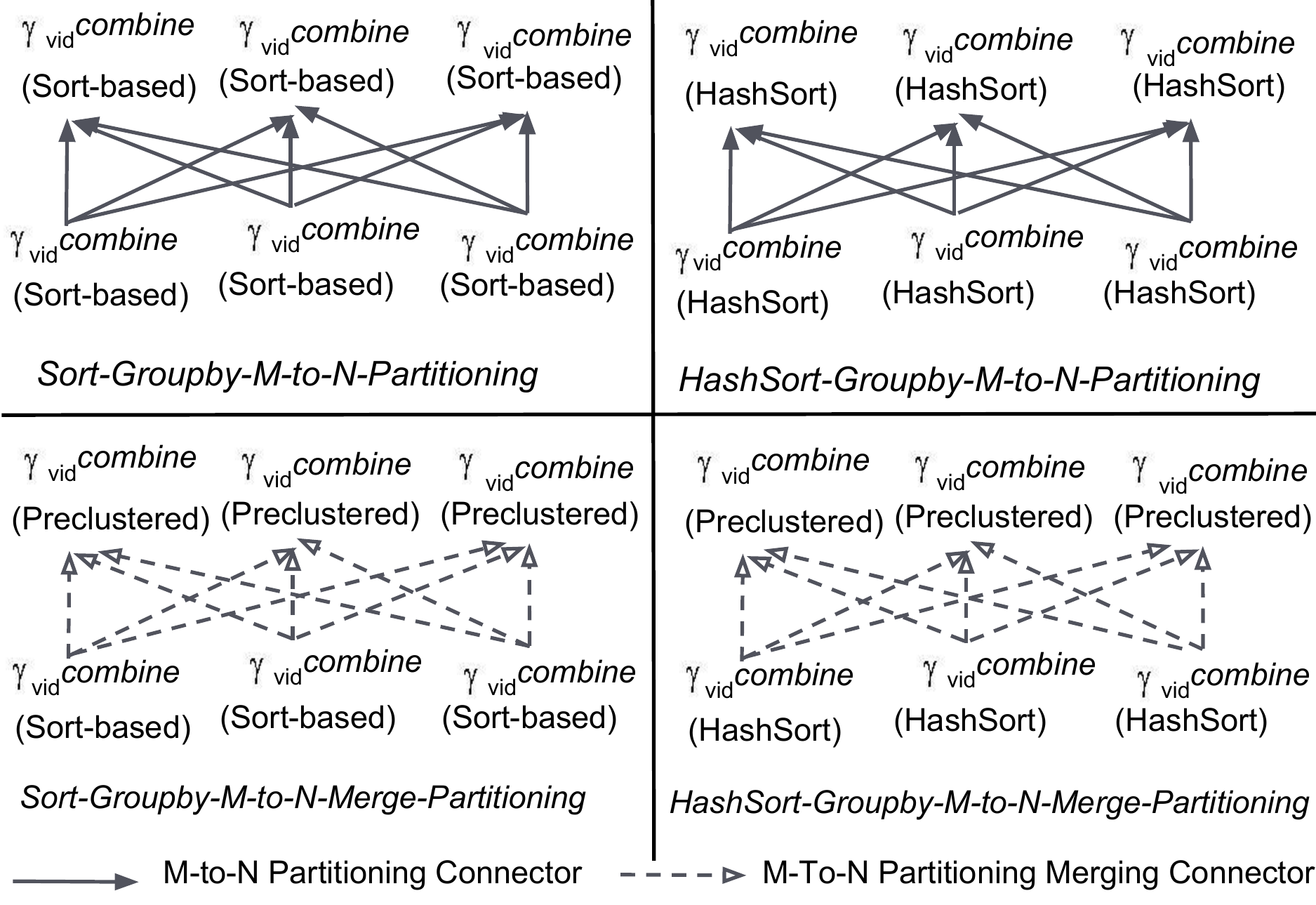}
    \vspace{-4ex}
    \caption{The four physical group-by strategies for the group-by operator which combines messages in Figure~\ref{fig:logicalplan}.}\label{fig:groups}
\end{figure}

Figure~\ref{fig:logicalplan} uses a logical group-by operator for message
combination.  For that, Pregelix leverages the three group-by operator
implementations mentioned in Section~\ref{hyracks}.  A preclustered group-by
can only be applied to input data that is already clustered by the grouping
key.  A HashSort group-by operator offers better performance (over sort-based
group-by) when the number of groups (the number of distinct message receivers
in our case) is small; otherwise, these two group-by operators perform
similarly.  In a parallel execution, the grouping is done by two stages---each
producer partitions its output (message) data by destination {\it vid}, and the
output is redistributed (according to
destination {\it vid}) to each consumer, which performs the final grouping
step.

Pregelix has four different parallel group-by strategies, as shown in
Figure~\ref{fig:groups}.  The lower two strategies use an m-to-n partitioning
merging connector and only need a simple one-pass pre-clustered group-by at
the receiver-side; however, in this case, receiver-side merging needs to
coordinate the input streams, which takes more time as the cluster size
grows. 
The upper two strategies use an m-to-n partitioning connector, which does not require such coordination; 
however, these strategies do not deliver sorted data streams to the receiver-side group-bys, so re-grouping is needed at the receiver-side.
 A fully pipelined policy is used
for the m-to-n partitioning connector in the upper two strategies, while in the
lower two strategies, a sender-side materializing pipelined policy is used by
the m-to-n partitioning merging connector to avoid possible deadlock scenarios
mentioned in the query scheduling literature~\cite{QE}.
The choice of which group-by strategy to use depends on the dataset, graph algorithm, and cluster.
We will further pursue this choice in our experiments (Section~\ref{experiments}).

\subsubsection{Message Delivery}\label{joins}

\begin{figure}[!t]
  \centering
    \includegraphics[width=\linewidth]{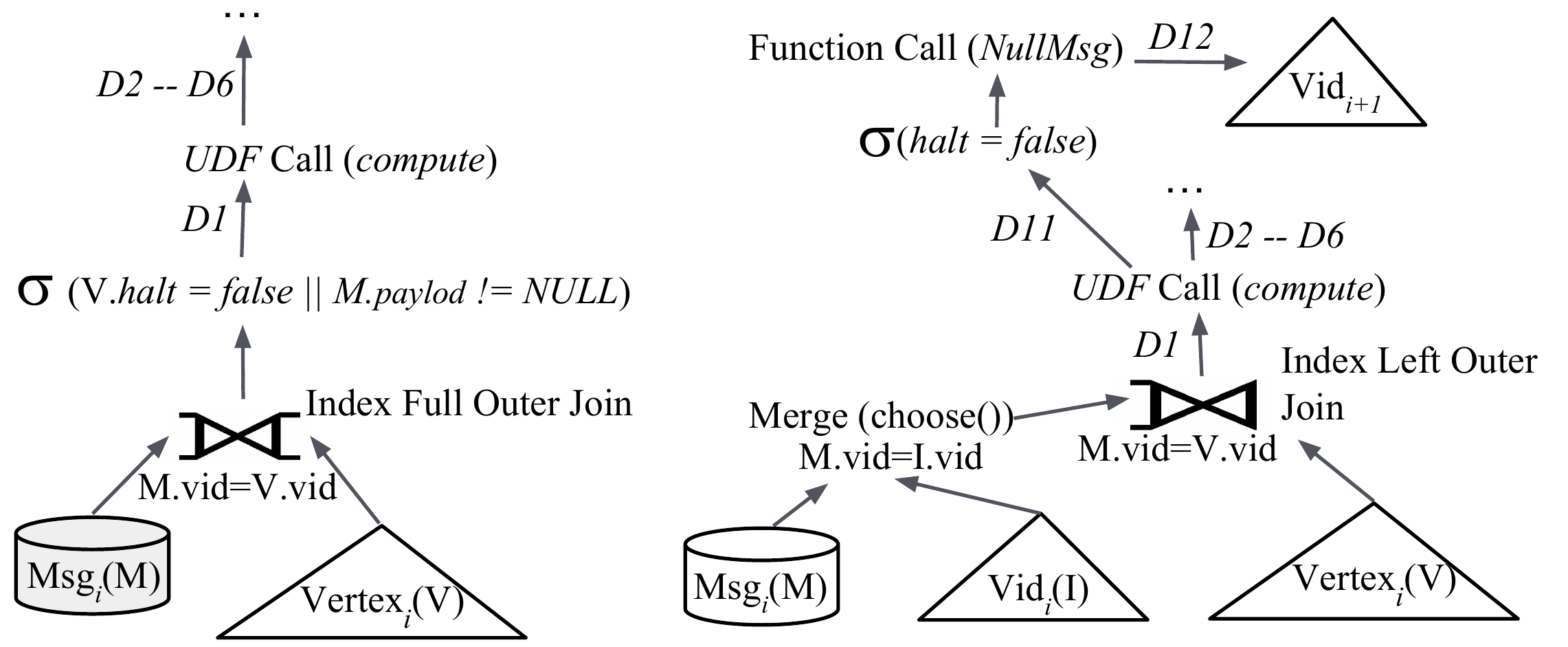}
    \vspace{-4ex}
    \caption{Two physical join strategies for forming the input to the \CodeIn{compute} UDF. 
    On the left is an index full outer join approach. On the right is an index left outer join approach.}\label{fig:joins}
\end{figure}

Recall that in Figure~\ref{fig:logicalplan}, a logical full-outer-join is used
to deliver messages to the right vertices and form the data needed to call the
\CodeIn{compute} UDF.  
For that, we use index-based joins because (a) vertices are already indexed by vid, and (b) 
all four group-by strategies in Figure~\ref{fig:groups} flow the (combined) messages out of their receiver-side 
group-bys in vid-sorted order, thereby producing vid-sorted Msg partitions.


Pregelix offers two physical choices for index-based joins---an index full
outer join approach and an index left outer join approach, as shown in
Figure~\ref{fig:joins}.  The full outer join plan scans the entire vertex index
to merge it with the (combined) messages.  This join strategy is suitable for
algorithms where most vertices are live (active) across supersteps (e.g., PageRank).
The left outer join plan prunes unnecessary vertex scans by first searching
the live vertex index for each (combined) incoming message, and it fits cases
where messages are sparse and only few vertices are live in every superstep
(e.g., single source shortest paths).  A user can control which join approach
Pregelix uses for a given job.  We now briefly explain the details of the two
join approaches.

{\bf Index Full Outer Join.} As shown in left side of Figure~\ref{fig:joins},
this plan is straightforwardly mapped from the logical plan.  The join operator
simply merges a partition of \CodeIn{Msg} and \CodeIn{Vertex} using a single pass.

{\bf Index Left Outer Join.} As shown in right of Figure~\ref{fig:joins},
this plan initially bulk loads another B-tree \CodeIn{Vid} with null messages
(vid, NULL) that are generated by a function \CodeIn{NullMsg}.  This index serves
to represent the set of currently active vertices.  The dataflows $D11$ and $D12$
in Figure~\ref{fig:joins} are (vid, halt) tuples and (vid, NULL) tuples
respectively.  Note that \CodeIn{Vid} is partitioned in the same way as
\CodeIn{Vertex}.  In the next superstep, a merge operator merges tuples from
\CodeIn{Msg} and \CodeIn{Vid} based on the equivalence of the \CodeIn{vid}
fields, and the {\it choose} function inside the operator selects tuples from
\CodeIn{Msg} to output when there are duplicates.  Output tuples of the merge
operator are sent to an index left outer join operator that probes the
\CodeIn{Vertex} index.  Tuples produced by the left outer join operator are
directly sent to the \CodeIn{compute} UDF.  The original filter operator
$\sigma$({\it V.halt=false$||$M.payload!= NULL}) in the logical plan is
logically transformed to the merge operator where tuples in \CodeIn{Vid}
satisfy {\it halt=false} and tuples in \CodeIn{Msg} satisfy {\it M.payload!=
NULL}.

To minimize data movements among operators, in a physical plan, we push the
filter operator, the UDF call of \CodeIn{compute}, the update to
\CodeIn{Vertex}, and the extraction (project) of fields in the output tuple of
\CodeIn{compute} into the upstream join operator as Hyracks ``mini-operators.''

\subsubsection{Global States and Graph Mutations}

To form the global halt state and aggregate state---see the two global
aggregations in Figure~\ref{fig:gs}---we leverage a standard two-stage
aggregation strategy.  Each worker pre-aggregates these state values (stage
one) and sends the result to a global aggregator that produces the final result
and writes it to HDFS (stage two).  The incrementing of superstep is also done by a trivial
dataflow.

The additions and removals of vertices in Figure~\ref{fig:addremove} are
applied to the \CodeIn{Vertex} relation by an index insert-delete operator.
For the group-by operator in Figure~\ref{fig:addremove}, we only do a
receiver-side group-by because the \CodeIn{resolve} function is not guaranteed
to be distributive and the connector for $D6$ (in Figure~\ref{fig:logicalplan})
is an m-to-n partitioning connector in the physical plan.

\subsubsection{Data Redistribution}\label{scheduling}

In a physical query plan, data redistribution is achieved by either the m-to-n
hash partitioning connector or the m-to-n hash partitioning merging connector
(mentioned in Section~\ref{hyracks}).  With the Hyracks provided
user-configurable scheduling, we let the location constraints of the join
operator (in Figure~\ref{fig:joins}) be the same as the places where partitions
of \CodeIn{Vertex} are stored across all the supersteps.
Also, the group-by operator (in Figure~\ref{fig:groups}) has the same location
constraints as the join operator, such that in all supersteps, \CodeIn{Msg} and
\CodeIn{Vertex} are partitioned in the same (sticky) way and the join between them
can be done without extra repartitioning.  Therefore, the only necessary data
redistributions in a superstep are (a) redistributing outgoing (combined)
messages from sender partitions to the right vertex partitions, and (b) sending
each vertex mutation to the right partition for addition or removal in the
graph data.

\subsection{Memory Management}\label{ooc}
Hyracks operators and access methods already provide support for out-of-core
computations.  The default Hyracks memory parameters work for all aggregated memory
sizes as long as there is sufficient disk space on the worker machines.  To support
both in-memory and out-of-core workloads, B-trees and LSM-trees both
leverage a buffer cache that caches partition pages and gracefully spills
to disk only when necessary using a standard replacement policy, i.e., LRU.
In the case of an LSM B-tree, some number of buffer pages are pinned in memory
to hold memory-resident B-tree components.  


The majority of the physical memory on a worker machine is divided into four parts:
the buffer cache for access methods of the \CodeIn{Vertex} relation; the
buffers for the group-by operator clones; the buffers for network channels; and
the file system cache for (a) temporary run files generated by group-by
operator clones, (b) temporary files for materialized data redistributions, and
(c) temporary files for the relation \CodeIn{Msg}.  The first three memory
components are explicitly controlled by Pregelix and can be tuned by a user,
while the last component is (implicitly) managed by the underlying OS.
Although the Hyracks runtime is written in Java, it uses a bloat-aware
design~\cite{ISMM} to avoid unnecessary memory bloat and to minimize the
performance impact of garbage collection in the JVM.

\subsection{Fault-Tolerance}~\label{ft}
Pregelix offers the same level of fault-tolerance as other Pregel-like
systems~\cite{Pregel, Giraph, Hama} by checkpointing states to HDFS at
user-selected superstep boundaries.  In our case, the states to be checkpointed
at the end of a superstep include \CodeIn{Vertex} and \CodeIn{Msg} (as well as
\CodeIn{Vid} if the left outer join approach is used).  The checkpointing of
\CodeIn{Msg} ensures that a user program does not need to be aware of failures.
Since \CodeIn{GS} stores its primary copy in HDFS, it need not be part of the
checkpoint.  A user job can determine whether or not to checkpoint after a superstep.
Once a node failure or disk failure happens, the failed machine is added into a
blacklist.

During recovery, Pregelix finds the latest checkpoint and reloads the states to
a newly selected set of failure-free worker machines.  Reloading states includes
two steps.  First, it kicks off physical query plans to scan, partition, sort,
and bulk load the entire \CodeIn{Vertex} and \CodeIn{Vid} (if any) from the
checkpoint into B-trees (or LSM B-trees), one per partition.  Second, it
executes another physical query plan to scan, partition, sort, and write the
checkpointed \CodeIn{Msg} data to each partition as a local file.

\subsection{Job Pipelining}\label{pipelining}
Pregelix can accept an array of jobs and pipeline between compatible contiguous
jobs without HDFS writes/reads nor index bulk-loads.  Two compatible jobs
should have a producer-consumer relationship regarding the output/input data
and share the same type of vertex---meaning, they interpret the corresponding
bits in the same way.  This feature was motivated by the genome
assembler~\cite{Velvet} application which runs six different graph cleaning
algorithms that are chained together for many rounds.  A user can choose to
enable this option to get improved performance with reduced fault-tolerance.

\subsection{Pregelix Software Components}\label{overview}
Pregelix supports the Pregel API introduced in Section~\ref{API} in Java, which
is very similar to the APIs of Giraph~\cite{Giraph} and Hama~\cite{Hama}.
Internally, Pregelix has a statistics collector, failure manager, scheduler,
and plan generator which run on a client machine after a job is submitted; it
also has a runtime context that stays on each worker machine.  We describe
each component below.

{\bf Statistics Collector.} The statistics collector periodically collects
statistics from the target Hyracks cluster, including system-wide counters such
as CPU load, memory consumption, I/O rate, network usage of each worker
machine, and the live machine set, as well as Pregel-specific statistics such
as the vertex count, live vertex count, edge count, and message count of a
submitted job.

{\bf Failure Manager.} The failure manager analyzes failure traces and recovers
from those failures that are indeed recoverable.  It only tries to recover from
interruption errors (e.g., a worker machine is powered off) and I/O related
failures (e.g., disk I/O errors); it just forwards application exceptions to
end users.  Recovery is done as mentioned in Section~\ref{ft}.

{\bf Scheduler.}  Based on the information obtained by the statistics collector and
the failure manager, the scheduler determines which worker machines to use to
run a given job.  The scheduler assigns as many partitions to a selected
machine as the number of its cores.  For each Pregel superstep, Pregelix sets
location constraints for operators in the manner mentioned in
Section~\ref{scheduling}.  For loading \CodeIn{Vertex} from HDFS~\cite{Hadoop},
the constraints of the data scanning operator (set by the scheduler) exploit
data locality for efficiency.

{\bf Plan Generator.} The plan generator generates physical query plans for
data loading, result writing, each single Pregel superstep, checkpointing, and
recovery.  The generated plan includes a physical operator DAG and a set of
location constraints for each operator.

{\bf Runtime Context.} The runtime context stores the cached \CodeIn{GS} tuple
and maintains the Pregelix-specific implementations of the Hyracks extensible
hooks to customize buffer, file, and index management.

\subsection{Discussion}\label{discussions}

Let us close this section by revisiting the issues
and opportunities presented in Section~\ref{problems}
and evaluating their implications in Pregelix:

\begin{list}{\labelitemi}{\leftmargin=1em}\itemsep 0pt \parskip 0pt

\item Out-of-core Support.  All the data processing operators as well as access
  methods we use have out-of-core support, which allows the physical query
  plans on top to be able to run disk-based workloads as well as multi-user
  workloads while retaining good in-memory processing performance.

\item Physical Flexibility.  The current physical choices spanning vertex
  storage (two options), message delivery (two alternatives), and message
  combination (four strategies) allow Pregelix to have sixteen ($2\times
  2\times 4$) tailored executions for different kinds of datasets, graph
  algorithms, and clusters.

\item Software Simplicity.  The implementations of all the described
  functionalities in this section leverage existing operator, connector, and access method libraries provided
  by Hyracks.  Pregelix does not involve modifications to the Hyracks runtime.

\end{list}

\section{Pregelix Case Studies}\label{usecases}
In this section, we briefly enumerate several Pregelix use cases, including
a built-in graph algorithm library, a study of graph connectivity problems,
and research on parallel genome assembly.

\begin{figure}[!t]
  \centering
    \includegraphics[width=0.95\linewidth]{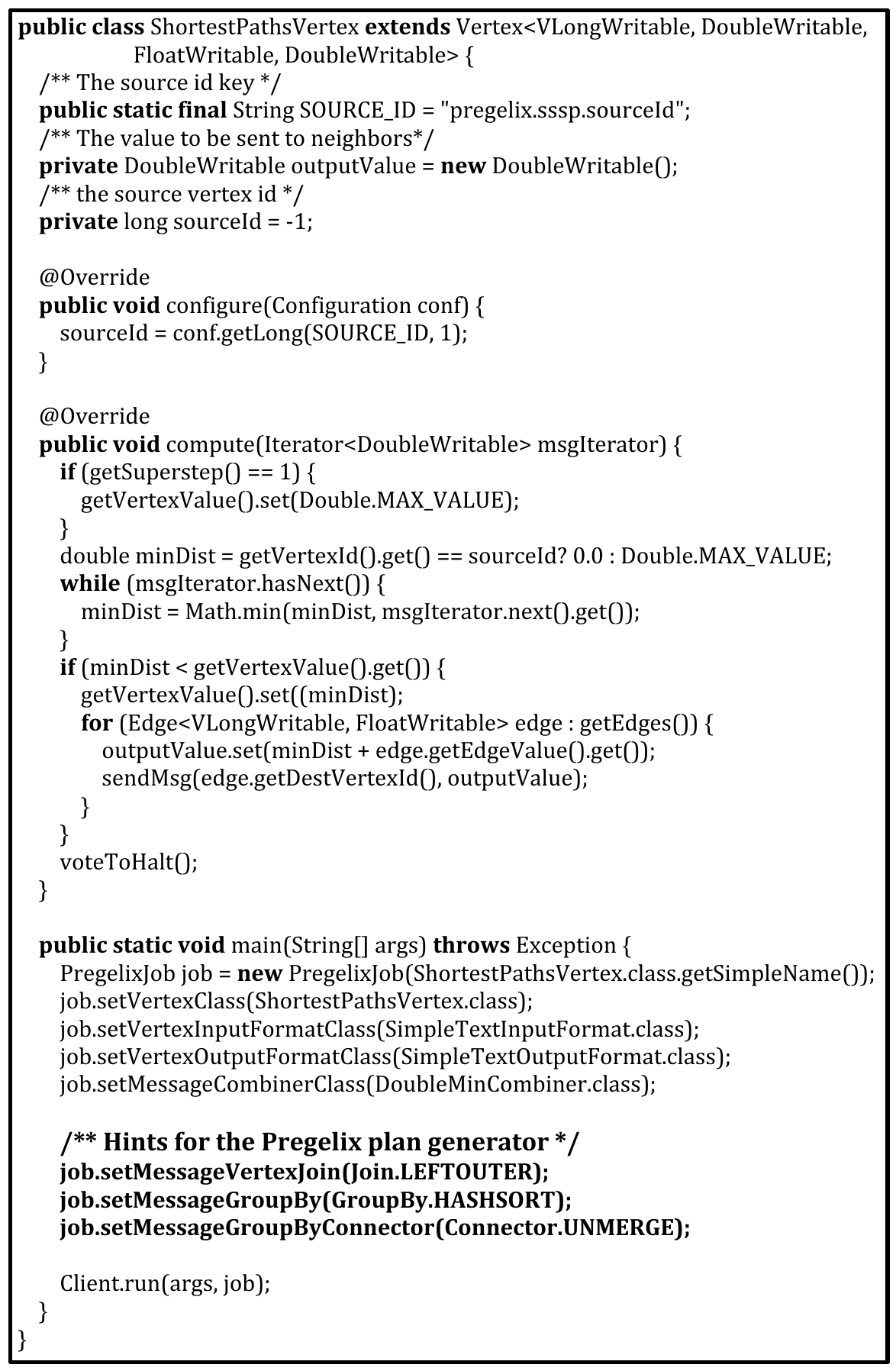}
    \vspace{-2ex}
    \caption{The implementation of the single source shortest paths algorithm on Pregelix.}\label{fig:code}
\end{figure}

{\bf The Pregelix Built-in Library.} The Pregelix software distribution comes
with a library that includes several graph algorithms such as PageRank, single
source shortest paths, connected components, reachability query, triangle
counting, maximal cliques, and random-walk-based graph sampling.
Figure~\ref{fig:code} shows the single source shortest paths implementation on
Pregelix, where hints for the join, group-by, and connector choices are set in
the \CodeIn{main} function.  Inside \CodeIn{compute}, the method calls to set
a vertex value and to send a message internally generate output tuples for the
corresponding dataflows.

{\bf Graph Connectivity Problems.} Using Pregelix, a graph analytics research
group in Hong Kong has implemented several graph algorithm building blocks such
as BFS (breath first search) spanning tree, Euler tour, list ranking, and
pre/post-ordering.  These building blocks have been used to develop advanced graph
algorithms such as bi-connected components for undirected graphs (e.g., road
networks) and strongly connected components for directed graphs (e.g., the
Twitter follower network)~\cite{Connectivity} . The group also scale-tested all
of their algorithms on a 60 machine cluster with 480 cores and 240 disks, using
Pregelix as the infrastructure.

{\bf Genome Assembly.} Genomix~\cite{Genomix} is a data-parallel genome
assembler built on top of Pregelix.  It first constructs a (very large) De
Bruijn graph~\cite{Velvet} from billions of genome reads, and then (a) cleans
the graph with a set of pre-defined subgraph patterns (described in
~\cite{Velvet}) and (b) merges available single paths into vertices iteratively
until all vertices can be merged to a single (gigantic) genome sequence.
Pregelix's support for the addition and removal of vertices is heavily used in
this use case.

\section{Experiments}\label{experiments}
This section compares Pregelix with several other popular parallel graph processing systems, 
including Giraph~\cite{Giraph}, Hama~\cite{Hama}, GraphLab~\cite{GraphLab}, and GraphX~\cite{GraphX}. Our comparisons cover execution time
(Section~\ref{exetime}), scalability (Section~\ref{scalability}), throughput
(Section~\ref{throughput}),  plan flexibility (Section~\ref{flexibility}), and software simplicity (Section~\ref{loc}).  
We conclude this section by summerizing our experimental results (Section~\ref{summary}).


\subsection{Experimental Setup}

We ran the experiments detailed here on a 32-node Linux IBM x3650 cluster with one
additional master machine of the same configuration. Nodes are connected with
a Gigabit Ethernet switch.  Each node has one Intel Xeon processor E5520
2.26GHz with four cores, 8GB of RAM, and two 1TB, 7.2K RPM hard disks.  

In our experiments, we leverage two real-world graph-based datasets.  The
first is the Webmap dataset~\cite{Webmap} taken from a crawl of the web in the
year $2002$.  The second is the BTC
dataset~\cite{BTC}, which is a
undirected semantic graph converted from the original Billion Triple Challenge 2009 RDF
dataset~\cite{BTCData}.  Table~\ref{tab:Webmap} (Webmap) and Table~\ref{tab:btc}
(BTC) show statistics for these graph datasets, including the full datasets as well as several down-samples and scale-ups\footnote{We
used a random walk graph sampler built on top of Pregelix to create scaled-down Webmap sample graphs of different sizes. 
To scale up the BTC data size, we deeply copied the original graph data
and renumbered the duplicate vertices with a new set of identifiers.}
that we use in our experiments.

Our evaluation examines the platforms' performance characteristics of three algorithms:
PageRank~\cite{PageRank}, SSSP (single source shortest paths)~\cite{GraphAlgorithm}, and CC (connected components)~\cite{GraphAlgorithm}.
On Pregelix and Giraph, the graph algorithms were coded in Java and all their source code
can be found in the Pregelix codebase\footnote{https://code.google.com/p/hyracks/source/browse/pregelix}.
The implementations of the three algorithms on Hama, GraphLab, and GraphX are
directly borrowed from their builtin examples.
The Pregelix default plan, which uses index full outer join, sort-based group-by,
an m-to-n hash partitioning connector, and B-tree vertex storage,
is used in Sections~\ref{exetime}, \ref{scalability}, and \ref{throughput}.
Pregelix's default maximum buffer cache size for access methods is set to $\frac{1}{4}$ the physical RAM size,
and its default maximum allowed buffer for each group-by operator instance is set to $64$MB. These two default
Pregelix memory settings are used in all the experiments.
For the local file system for Pregelix, we use the ext3 file system; for the distributed file system, we use HDFS version $1.0.4$. 
In all experiments, we use Giraph trunk version (revision 770), 
Hama version 0.6.4, GraphLab version 2.2 (PowerGraph), and Spark~\cite{Spark} version 0.9.1 for GraphX.
Our GraphLab setting has been confirmed by its primary author.
We tried our best to let each system use all the CPU cores and all available RAM on each worker machine.


\begin{table}[!t]
\begin{center}
\scriptsize
\begin{tabular}{|c||r|r|r|r|}
\cline{1-5}
           \textbf{Name} & \textbf{Size} &\textbf{\#Vertices} &\textbf{\#Edges} & \textbf{Avg. Degree}  \\
\hline     	Large & 71.82GB & 1,413,511,390 & 8,050,112,169 &  5.69\\
\cline{1-5}     Medium & 31.78GB & 709,673,622 & 2,947,603,924 & 4.15 \\
\cline{1-5}     Small & 14.05GB & 143,060,913 & 1,470,129,872  & 10.27\\
\cline{1-5}     X-Small & 9.99GB & 75,605,388 & 1,082,093,483 & 14.31\\
\cline{1-5}     Tiny & 2.93GB & 25,370,077 & 318,823,779 & 12.02 \\
\cline{1-5}
\end{tabular}
\end{center}
\vspace{-3ex}
\caption{The Webmap dataset (Large) and its samples.} \label{tab:Webmap}
\end{table}

\begin{table}[!t]
\begin{center}
\scriptsize
\begin{tabular}{|c||r|r|r|r|}
\cline{1-5}
           \textbf{Name} & \textbf{Size} &\textbf{\#Vertices} &\textbf{\#Edges} & \textbf{Avg. Degree}  \\
\hline     	Large & 66.48GB & 690,621,916 & 6,177,086,016 & 8.94\\
\cline{1-5}     Medium & 49.86GB & 517,966,437 & 4,632,814,512 & 8.94\\
\cline{1-5}     Small & 33.24GB & 345,310,958 & 3,088,543,008 & 8.94 \\
\cline{1-5}     X-Small & 16.62GB & 172,655,479 & 1,544,271,504 & 8.94\\
\cline{1-5}     Tiny & 7.04GB & 107,706,280 & 607,509,766 & 5.64\\
\cline{1-5}
\end{tabular}
\end{center}
\vspace{-3ex}
\caption{The BTC dataset (X-Small) and its samples/scale-ups.} \label{tab:btc}
\end{table}

\begin{figure*}[!t]
\begin{center}
\hspace*{-2ex}
\begin{tabular}{ccc}
\begin{minipage}[t]{0.33\linewidth}
\includegraphics[width=0.95\columnwidth]{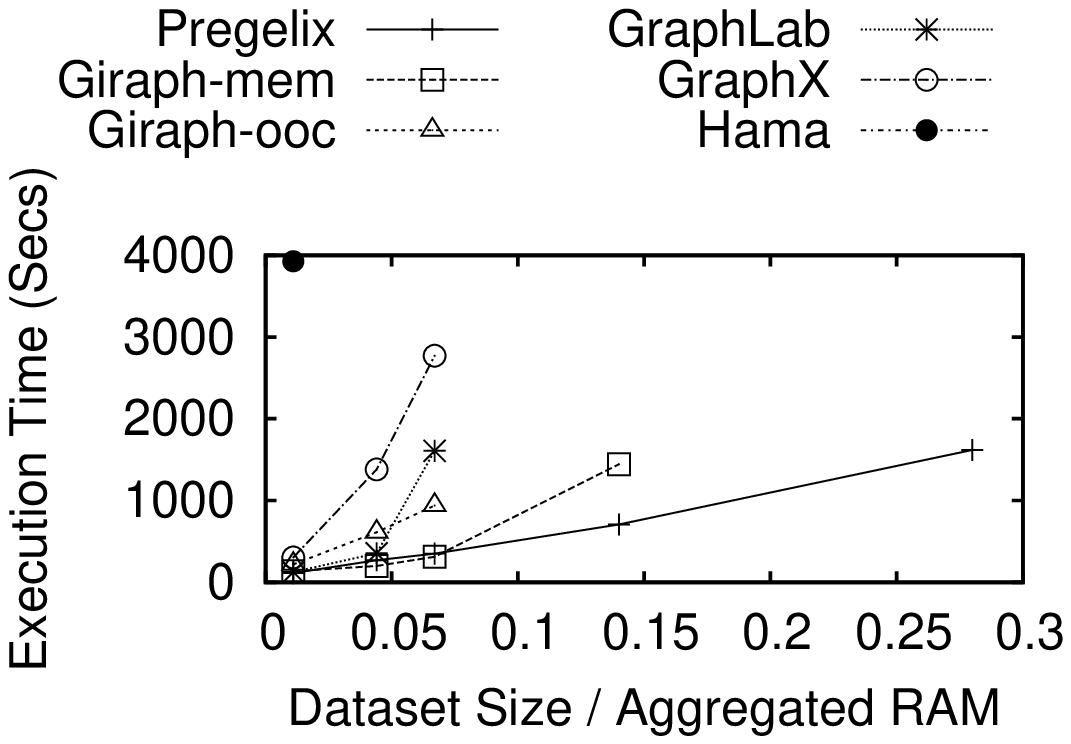}
\end{minipage}
&
\begin{minipage}[t]{0.33\linewidth}
\includegraphics[width=0.95\columnwidth]{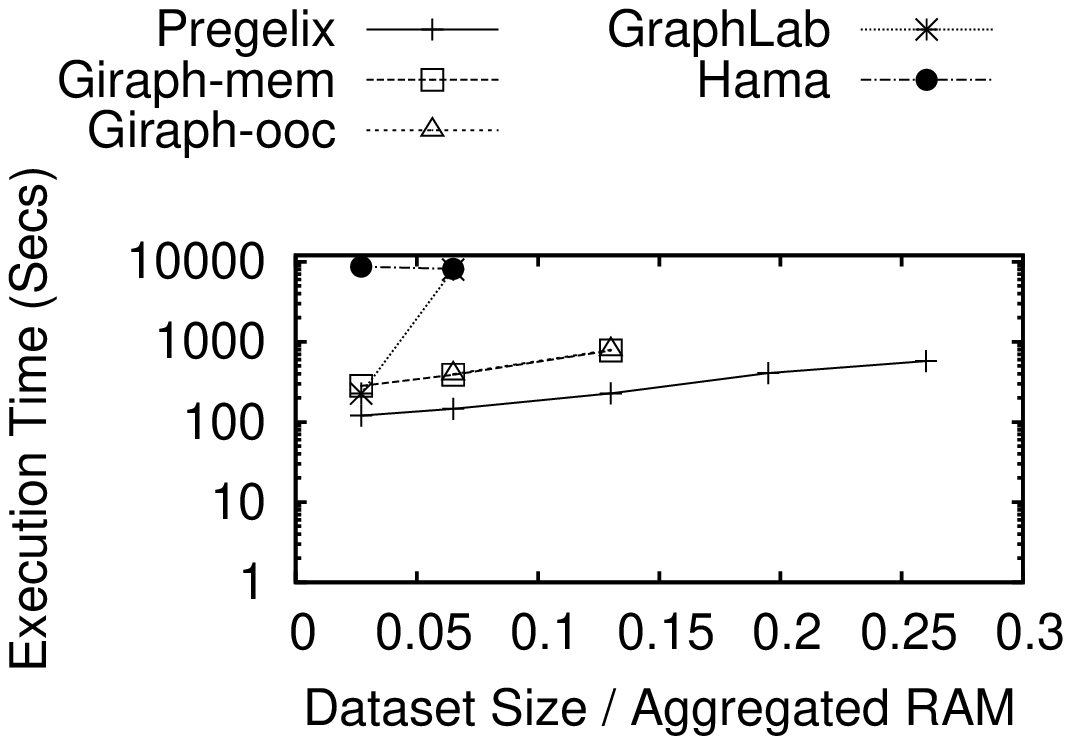}
\end{minipage}
&
\begin{minipage}[t]{0.33\linewidth}
\includegraphics[width=0.95\columnwidth]{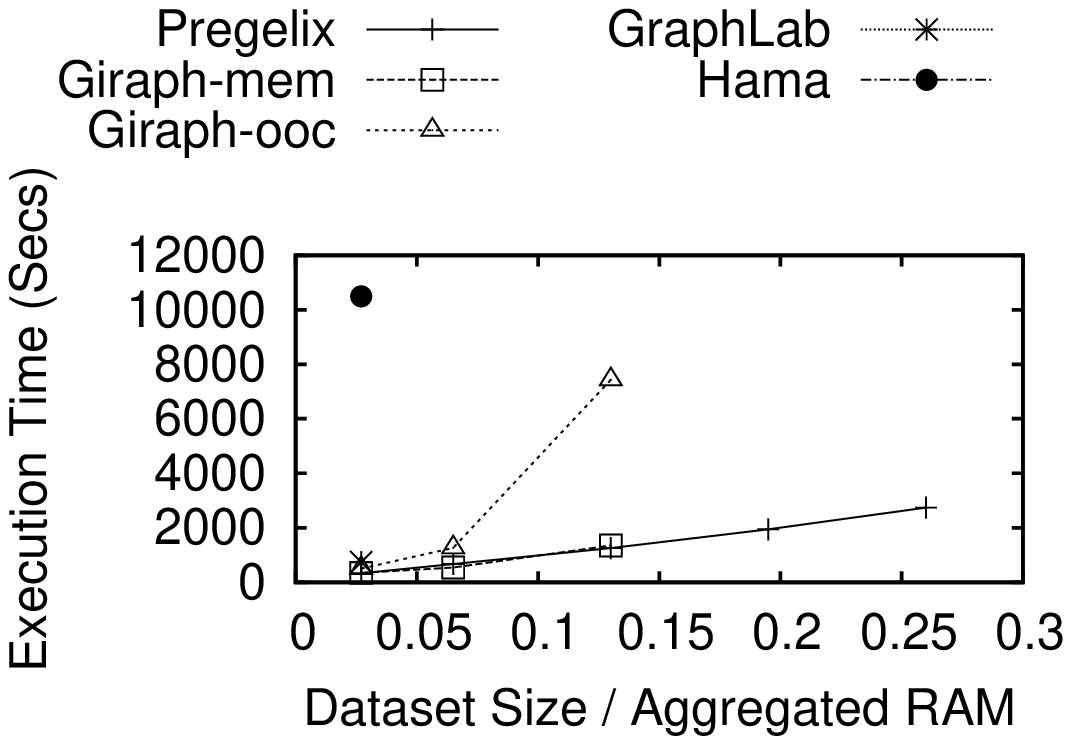}
\end{minipage}\\
(a) PageRank for Webmap datasets &(b) SSSP for BTC datasets &(c) CC for BTC datasets\\
\end{tabular}
\vspace{-2ex}
 \caption{Overall execution time (32-machine cluster). Neither Giraph-mem nor Giraph-ooc can work properly when the ratio of dataset size to the aggregated RAM size exceeds 0.15;
GraphLab starts failing when the ratio of dataset size to the aggregated RAM size exceeds 0.07;  Hama fails on even smaller datasets; GraphX fails to load the smallest BTC dataset sample BTC-Tiny.} \label{fig:response32}
\end{center}
\end{figure*}

\begin{figure*}[!t]
\begin{center}
\hspace*{-2ex}
\begin{tabular}{ccc}
\begin{minipage}[t]{0.33\linewidth}
\includegraphics[width=0.95\columnwidth]{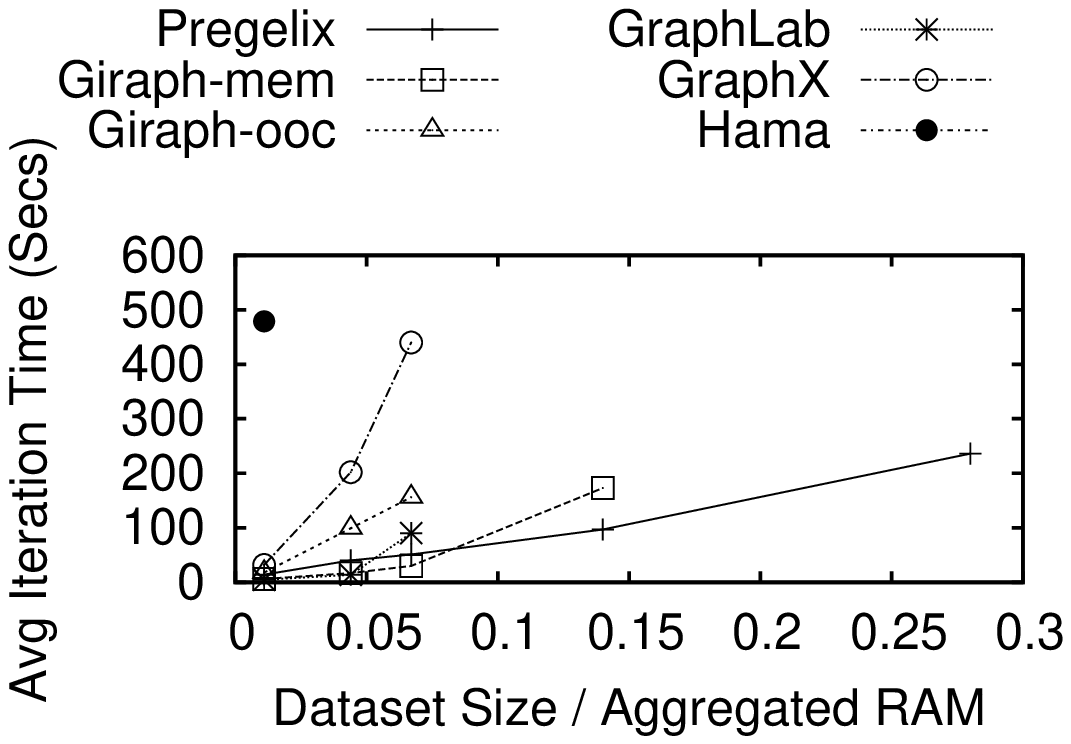}
\end{minipage}
&
\begin{minipage}[t]{0.33\linewidth}
\includegraphics[width=0.95\columnwidth]{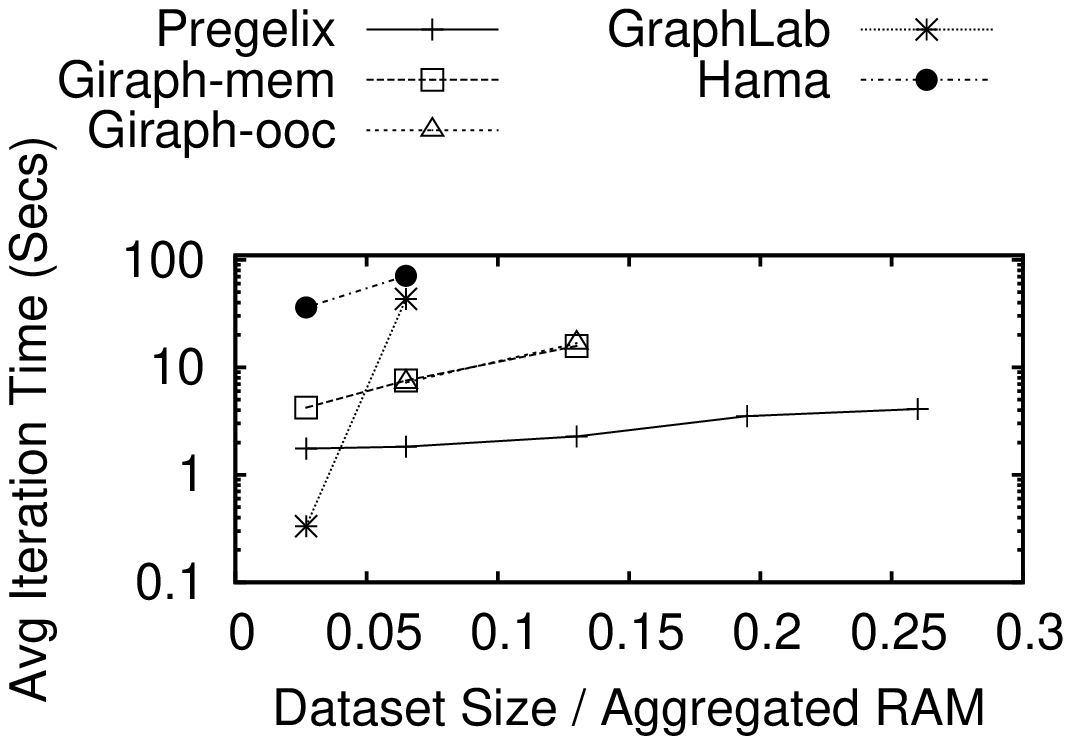}
\end{minipage}
&
\begin{minipage}[t]{0.33\linewidth}
\includegraphics[width=0.95\columnwidth]{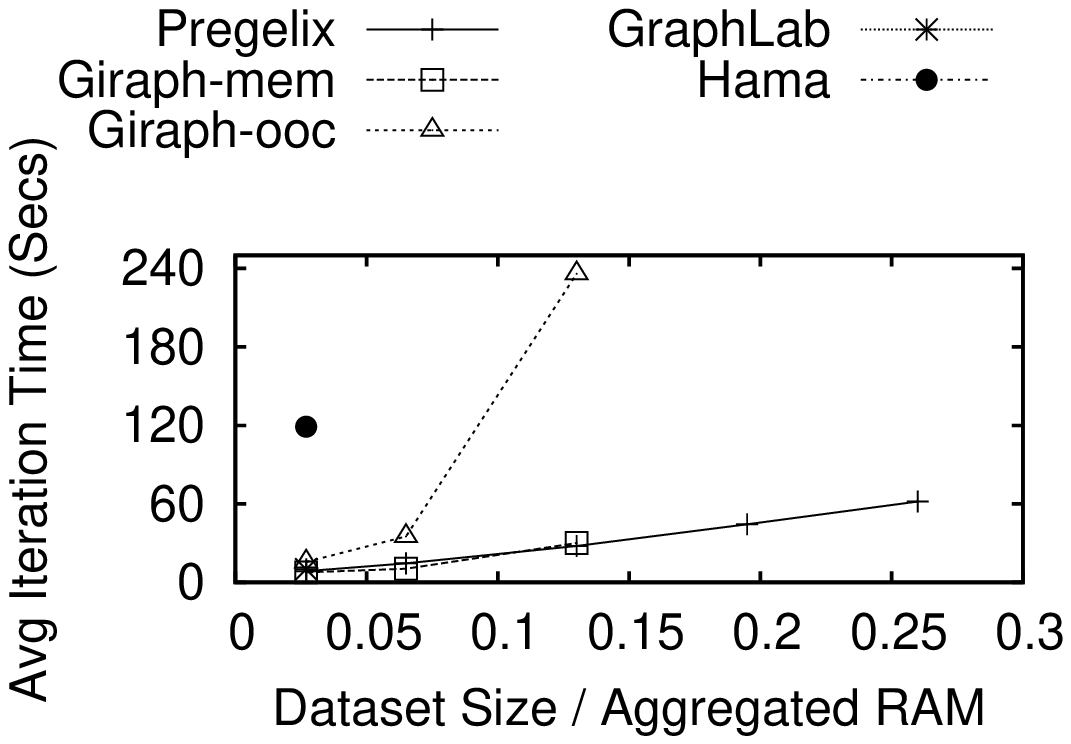}
\end{minipage}\\
(a) PageRank for Webmap datasets &(b) SSSP for BTC datasets & (c) CC for BTC datasets\\
\end{tabular}
\vspace{-2ex}
 \caption{Average iteration execution time (32-machine cluster).} \label{fig:iteration32}
\vspace{-2ex}
\end{center}
\end{figure*}

\subsection{Execution Time}~\label{exetime}
In this experiment, we evaluate the execution times of all systems by running each of
the three algorithms on the 32-machine cluster.  As the input data for PageRank we use the
Webmap dataset because PageRank is designed for ranking web pages,
and for the SSSP and CC algorithms we use the BTC dataset.  Since a Giraph
user needs to explicitly specify apriori whether a job is in-memory or out-of-core, we
measure both of these settings (labeled Giraph-mem and Giraph-ooc,
respectively) for Giraph jobs regardless of the RAM size.

Figure~\ref{fig:response32} plots the resulting {\it overall} Pregel job execution times for the
different sized datasets, and Figure~\ref{fig:iteration32} shows the average
{\it per-iteration} execution time for all iterations.  In both figures, the x-axis is the input dataset size
relative to the cluster's aggregated RAM size, and the y-axis is the execution time.  (Note
that the volume of exchanged messages can exhaust memory even if the initial input graph dataset can
fit into memory.)  

Figure~\ref{fig:response32} and Figure~\ref{fig:iteration32} 
show that while Pregelix scales to out-of-core workloads, Giraph fails to run
the three algorithms once the relative dataset size exceeds $0.15$, even with its out-of-core
setting enabled. 
Figures~\ref{fig:response32}(a)(c) and \ref{fig:iteration32}(a)(c) show
that when the computation has sufficient memory, Pregelix offers comparable
execution time to Giraph for message-intensive workloads such as PageRank and CC. For PageRank, Pregelix runs up to 2$\times$ slower on 
very small datasets but up to 2$\times$ faster on large datasets than Giraph. 
For CC, both systems perform similarly fast for all cases.
Figure~\ref{fig:response32}(b) and
Figure~\ref{fig:iteration32}(b) further demonstrate that the Pregelix default plan offers
$3.5\times$ overall speedup and $7\times$ per-iteration speedup over Giraph for
message-sparse workloads like SSSP even for relatively small datasets.  
All sub-figures in Figure~\ref{fig:response32} and Figure~\ref{fig:iteration32} 
show that for in-memory workloads (when the relative dataset size is less than $0.15$), Giraph has steeper (worse) size-scaling curves than Pregelix. 

Compared to Giraph,
GraphLab, GraphX, and Hama start failing on even smaller datasets, with even steeper size-scaling curves. 
GraphLab has the best average per-iteration execution time on small datasets (e.g., up to 5$\times$ faster than Pregelix and up to 12$\times$ 
faster than Giraph, on BTC-Tiny), but performs worse than Giraph and Pregelix on larger datasets (e.g.,
up to 24$\times$ slower than Pregelix and up to 6$\times$ slower than Giraph, on BTC-X-Small).
GraphX cannot successfully load the BTC-Tiny dataset on the 32-machine cluster, therefore
its results for SSSP and CC are missing.   



\subsection{System Scalability}\label{scalability}
\begin{figure*}[!t]
\begin{center}
\hspace*{-2ex}
\begin{tabular}{ccc}
\begin{minipage}[t]{0.33\linewidth}
\includegraphics[width=0.90\columnwidth]{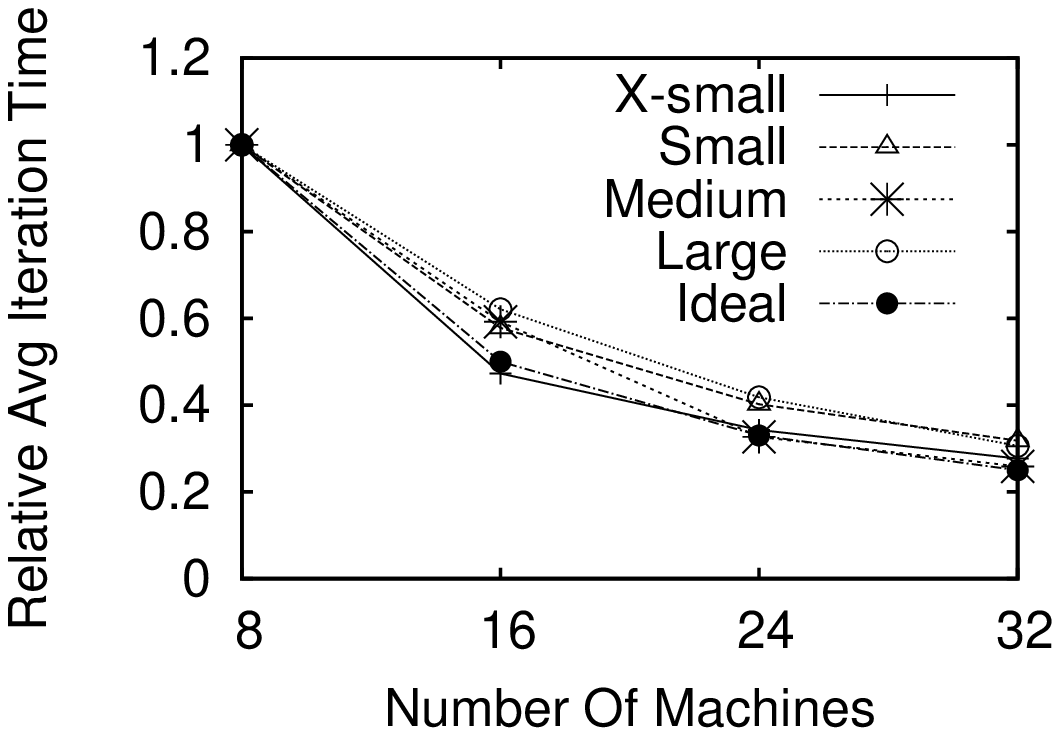}
\end{minipage}
&
\begin{minipage}[t]{0.33\linewidth}
\includegraphics[width=0.90\columnwidth]{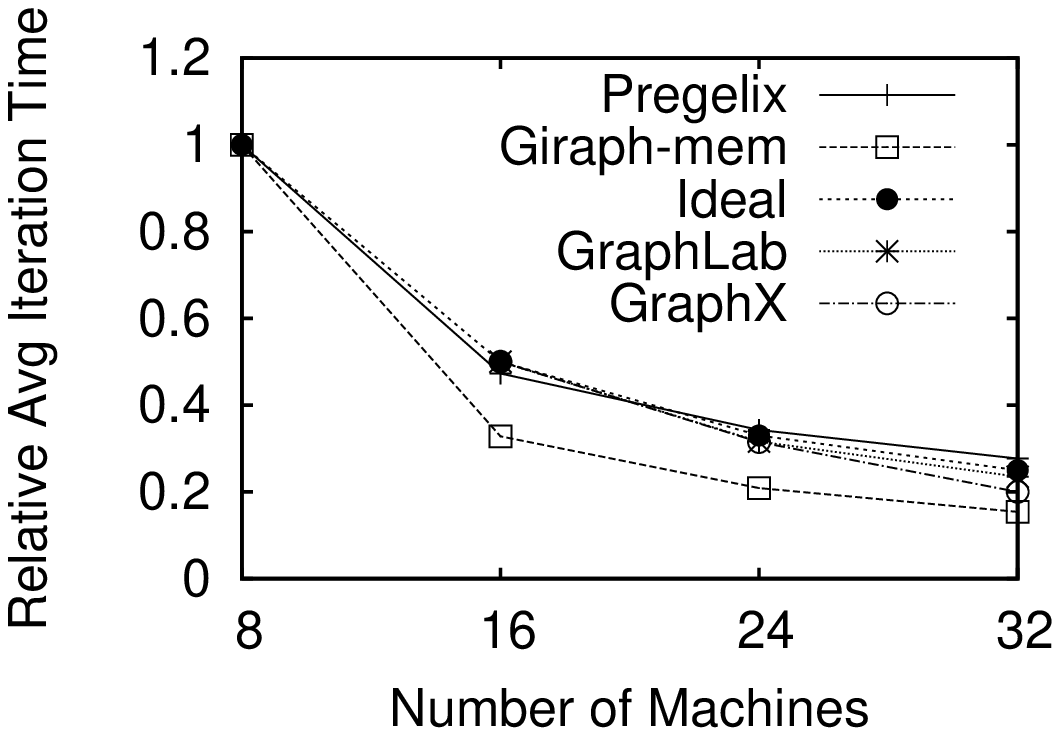}
\end{minipage}
&
\begin{minipage}[t]{0.33\linewidth}
\includegraphics[width=0.90\columnwidth]{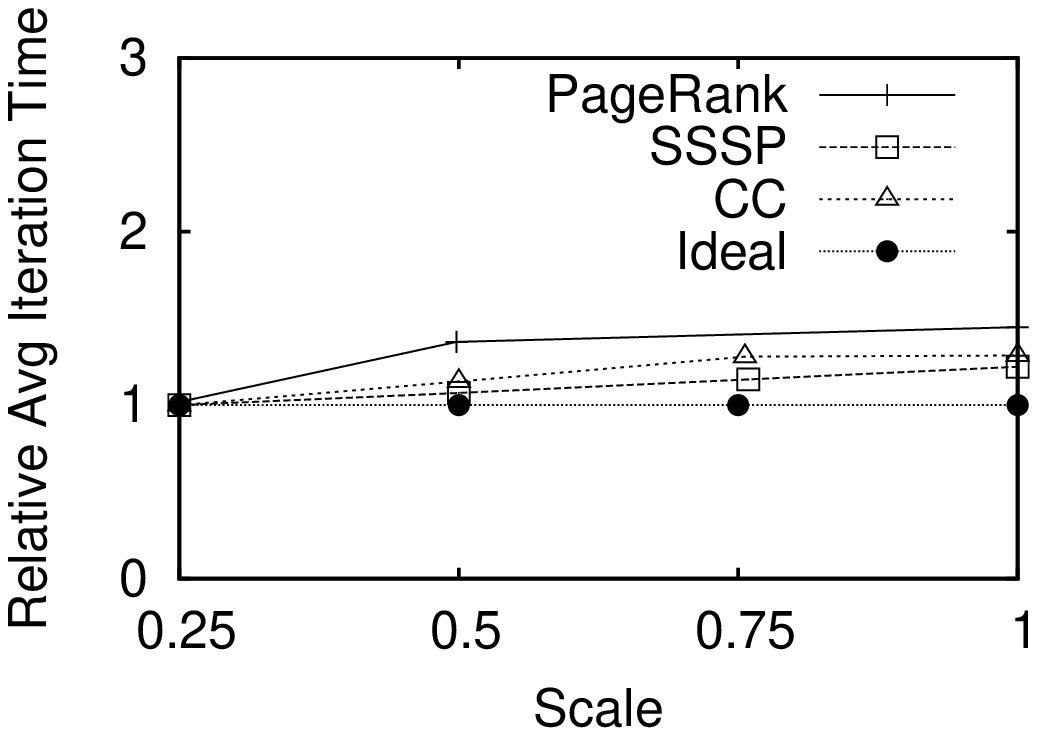}
\end{minipage}\\
(a) Pregelix Speedup (PageRank) &(b) Speedup (PageRank) for Webmap-X-Small & (c) Pregelix Scaleup (PageRank, SSSP, CC)\\
\end{tabular}
\vspace{-2ex}
 \caption{Scalability  (run on 8-machine, 16-machine, 24-machine, 32-machine clusters).} \label{fig:scale}
\vspace{-2ex}
\end{center}
\end{figure*}

\begin{figure*}[!t]
\begin{center}
\hspace*{-2ex}
\begin{tabularx}{\linewidth}{XXXX}
\begin{minipage}[t]{\linewidth}
\includegraphics[width=\columnwidth]{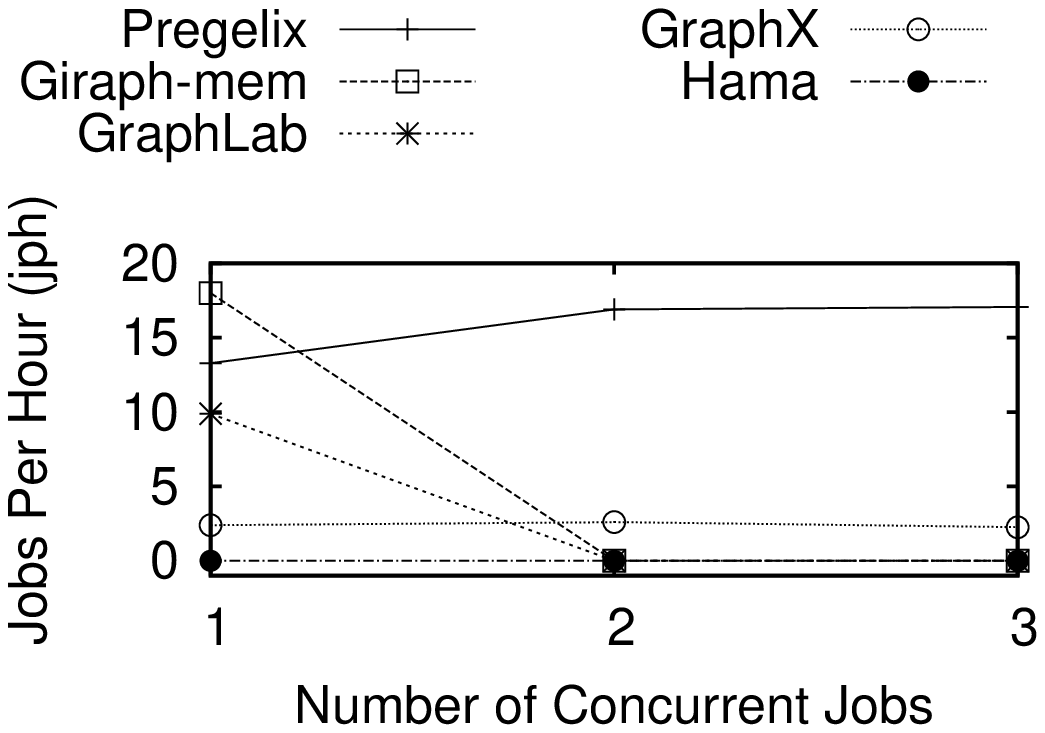}
\end{minipage}
&
\begin{minipage}[t]{\linewidth}
\includegraphics[width=\columnwidth]{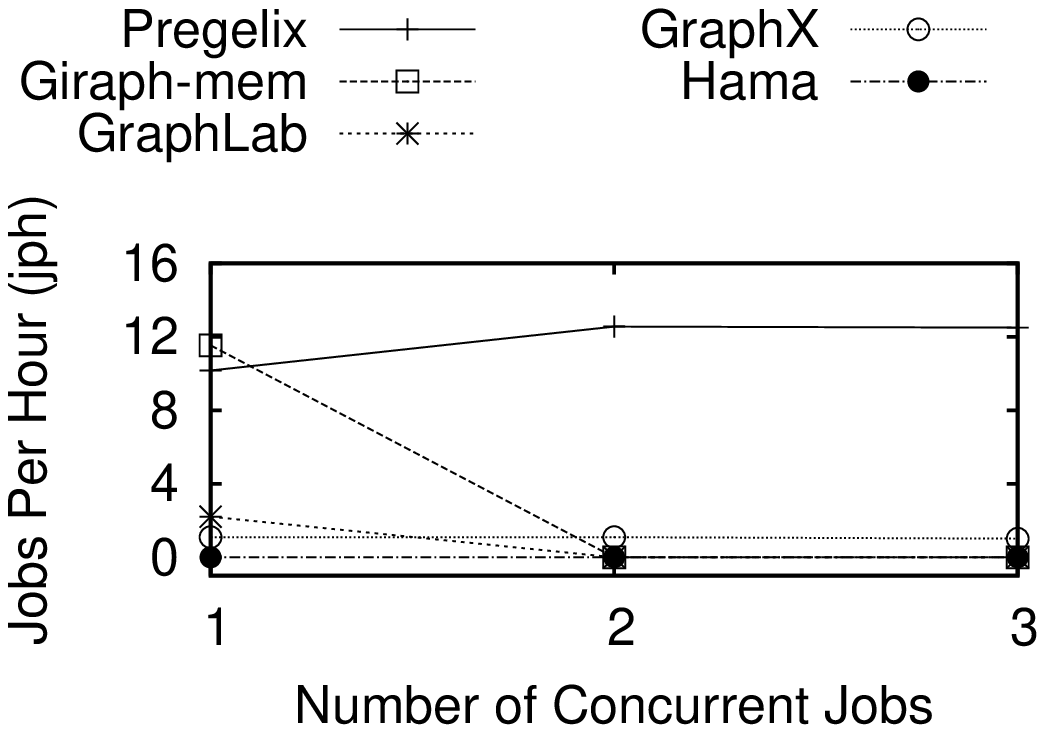}
\end{minipage}
&
\begin{minipage}[t]{\linewidth}
\includegraphics[width=\columnwidth]{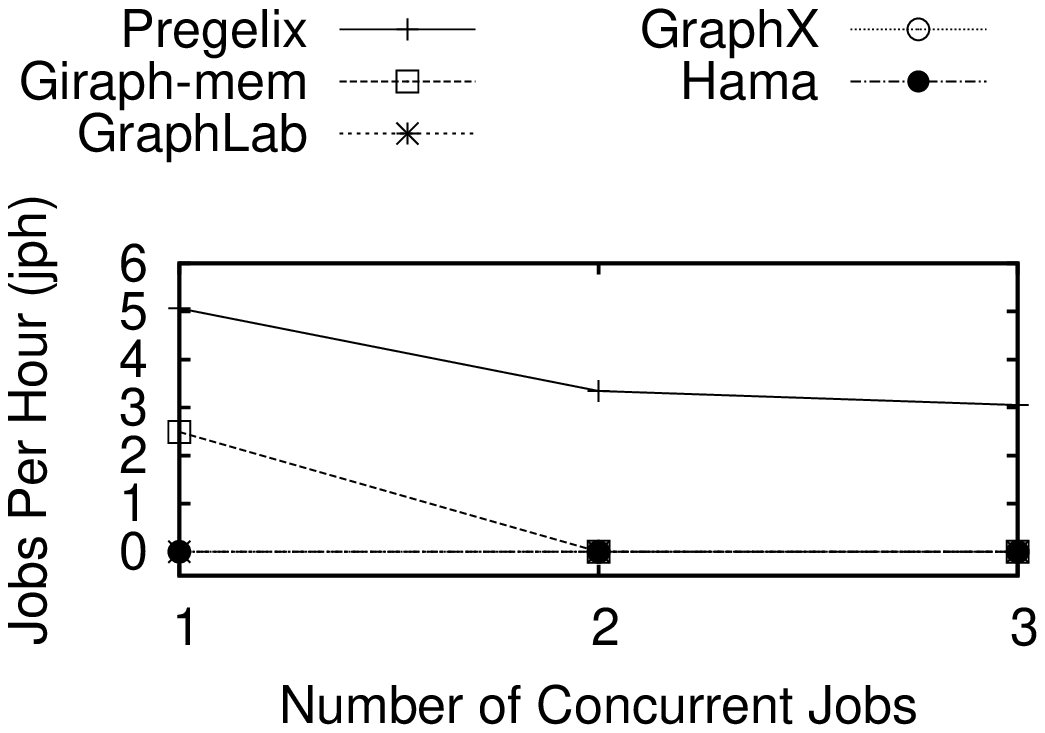}
\end{minipage}
&
\begin{minipage}[t]{\linewidth}
\includegraphics[width=\columnwidth]{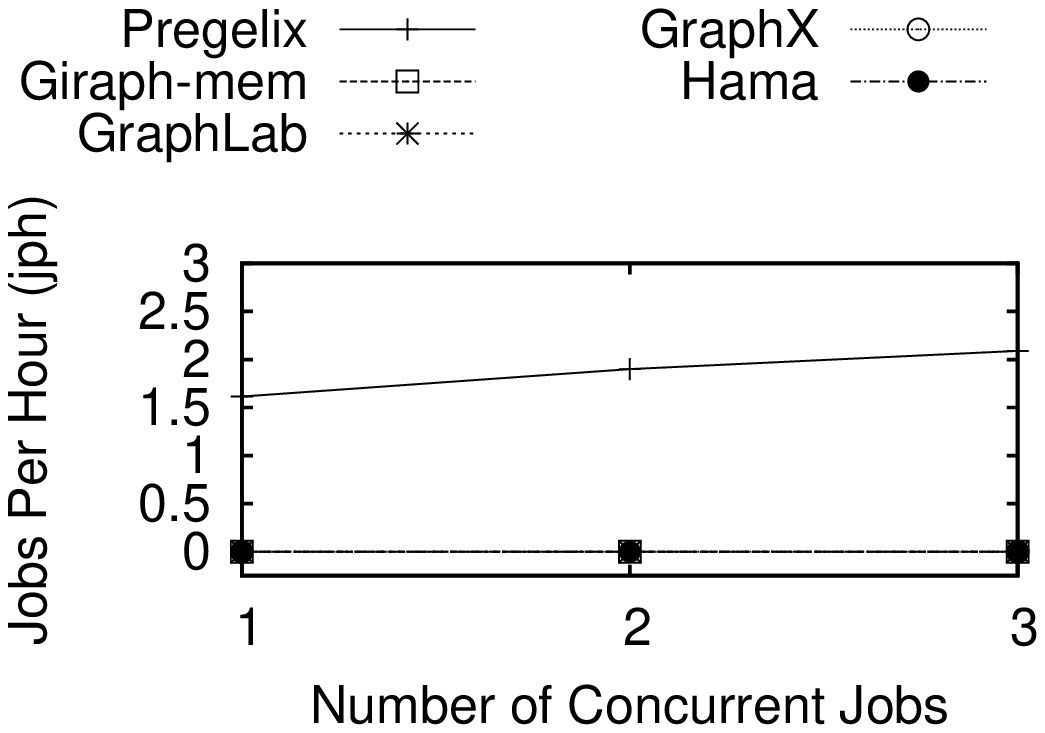}
\end{minipage}\\
(a)  Webmap-X-Small (always in-memory)  &(b) Webmap-Small (in-memory to minor disk usage) & (c) Webmap-Medium (in-memory to disk-based) & (d) Webmap-Large (always disk-based)\\
\end{tabularx}
\vspace{-1ex}
 \caption{Throughput (multiple PageRank jobs are executed in the 32-machine cluster with different sized datasets).} \label{fig:throughput}
\vspace{-2ex}
\end{center}
\end{figure*}

Our system scalability experiments run the different systems on varying sized
clusters for each of the different dataset sizes.  Figure~\ref{fig:scale}(a)
plots the parallel speedup for PageRank on Pregelix going from
8 machines to 32 machines.  The x-axis is the number of machines,
and the y-axis is the average per-iteration execution time relative to the time
on 8 machines.  As the number of machines increases, the message combiner for
PageRank becomes less effective and hence the total volume of data transferred
through the network gets larger, though the CPU load of each individual machine
drops.  Therefore, in Figure~\ref{fig:scale}(a), the parallel speedups are
close to but slightly worse than the ``ideal" case in which there are no message
overheads among machines.  For the other systems, the PageRank implementations
for GraphLab and GraphX fail to run Webmap samples beyond the 9.99GB case
when the number of machines is 16; Giraph has the same issue when the number of
machines is 8.  Thus, we were only able to compare the parallel PageRank speedups
of Giraph, GraphLab, GraphX, and Pregelix with the Webmap-X-Small dataset.
The results of this small data case are in Figure~\ref{fig:scale}(b); Hama is not included
because it cannot run even the Webmap-X-Small dataset on our cluster configurations. 
The parallel speedup of Pregelix is very close to the ideal line, while Giraph, GraphLab,
and GraphX exhibit even better speedups than the ideal.
The apparent super-linear parallel speedups of Giraph, GraphLab, and GraphX are consistent with
the fact that they all perform super-linearly worse when the volume of data assigned to a slave
machine increases, as can be seen in Figures~\ref{fig:response32} and \ref{fig:iteration32}.

Figure~\ref{fig:scale}(c) shows the parallel scale up of the three algorithms
for Pregelix. Giraph, GraphLab, GraphX, and Hama results are not shown because they could not run all these cases.  In this figure, the x-axis is the ratio of the sampled (or scaled)
dataset size over the largest (Webmap-Large or BTC-Large) dataset size. The number of machines is proportional
to this ratio and 32 machines are used for scale 1.0. The y-axis is the
average per-iteration execution time relative to the time at the smallest scale.  In the
ideal case, the y-axis value would stay at 1.0 for all the scales, while in reality,
the three Pregel graph algorithms all have network communication and thus cannot achieve the
ideal.  The SSSP algorithm sends fewer messages than the other two algorithms, so
it is the closest to the ideal case.

\subsection{Throughput}\label{throughput}
In the current version of GraphLab, Hama, and Pregelix, each submitted job is executed immediately, regardless of the current activity 
level on the cluster.
Giraph leverages Hadoop's admission control; for the purpose of testing concurrent workloads, we let the number of Hadoop map task slots in each task tracker be 3.
GraphX leverages the admission control of Spark, such that jobs will be executed sequentially if available resources cannot meet the overall
requirements of concurrent jobs.
In this experiment, we compare the job throughput of all the systems by submitting jobs concurrently.
We ran PageRank jobs on the 32-machine cluster using four different samples of the Webmap dataset (X-Small, Small, Medium, and Large)
with various levels of job concurrency.  Figure~\ref{fig:throughput} reports
how the number of completed jobs per hour (\textbf{jph}) changes with the number of concurrent jobs.
The results for the four Webmap samples represent four different cases respectively:
 
\begin{list}{\labelitemi}{\leftmargin=1em}\itemsep 0pt \parskip 0pt
\item Figure~\ref{fig:throughput}(a) uses Webmap-X-Small. Moving
from serial job execution to concurrent job execution, data processing
remains in-memory but the CPU resources go from dedicated to shared.
In this case, Pregelix achieves a higher jph when there are two or three concurrent jobs
than when job execution is serial. 

\item Figure~\ref{fig:throughput}(b) uses Webmap-Small. In this case, serial job execution
does in-memory processing, but concurrent job execution introduces a small amount of disk I/O due to spilling.
When two jobs run concurrently, each job incurs about 1GB of I/O;
when three jobs run concurrently, each does about 2.7GB of I/O.
In this situation, still, the Pregelix jph is higher in concurrent execution than in serial execution.

\item Figure~\ref{fig:throughput}(c) uses Webmap-Medium. In this case,
serial job execution allows for in-memory processing, but allowing concurrent execution
exhausts memory and causes a significant amount of I/O for each individual job.
For example, when two jobs run concurrently, each job incurs about 10GB of I/O;
when three jobs run concurrently, each job does about 27GB of I/O. 
In this case,
jph drops significantly at the boundary where I/O significantly comes into the picture. 
The point with two concurrent jobs is such a boundary for Pregelix in Figure~\ref{fig:throughput}(c).

\item Figure~\ref{fig:throughput}(d) uses the full Webmap (Webmap-Large). In this case, processing is always disk-based regardless of the concurrency level.
For this case, Pregelix jph once again increases with the increased level of concurrency; 
this is because the CPU utilization is increased (by about 20\% to 30\%) with added concurrency.

\end{list}  

These results suggest that it would be worthwhile to develop intelligent job admission control policies
to make sure that Pregelix
runs with the highest possible throughput all the time in our future work.
In our experiments, the Spark scheduler for GraphX always runs concurrent jobs sequentially due to the lack of memory and CPU resources.
Giraph, GraphLab, and Hama all failed to support concurrent jobs in our experiments for all four
cases due to limitations regarding
memory management and out-of-core support; they each need additional work to operate in this region.

\begin{figure*}[!t]
\begin{center}
\hspace*{-2ex}
\begin{tabular}{ccc}
\begin{minipage}[t]{0.33\linewidth}
\includegraphics[width=0.90\columnwidth]{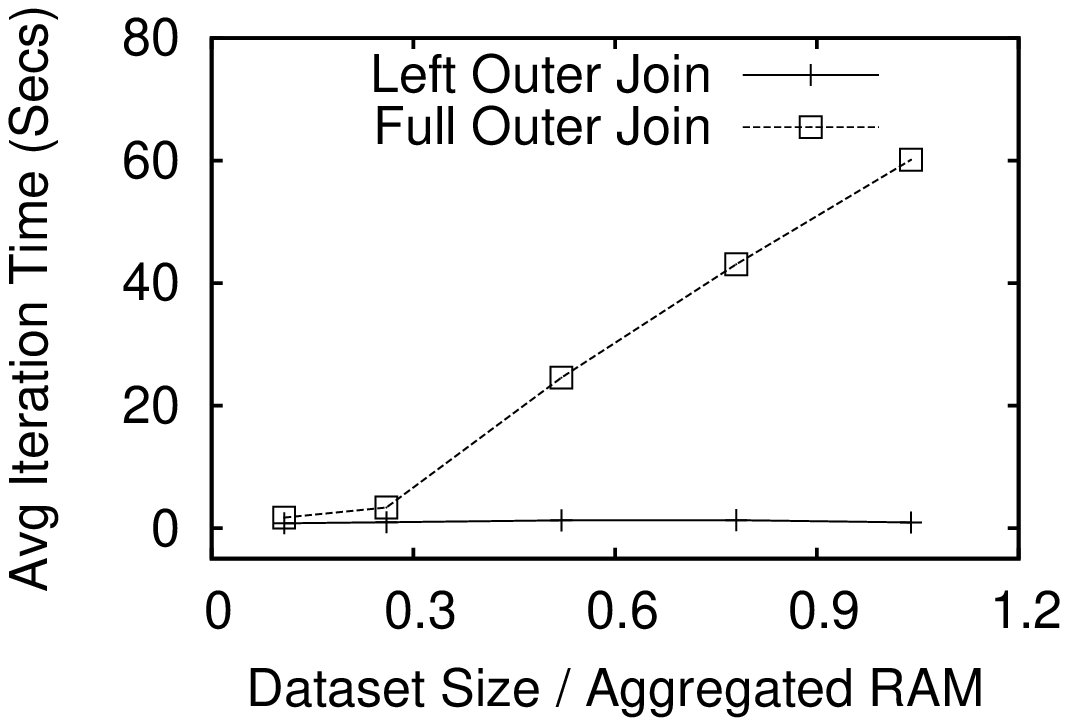}
\end{minipage}
&
\begin{minipage}[t]{0.33\linewidth}
\includegraphics[width=0.90\columnwidth]{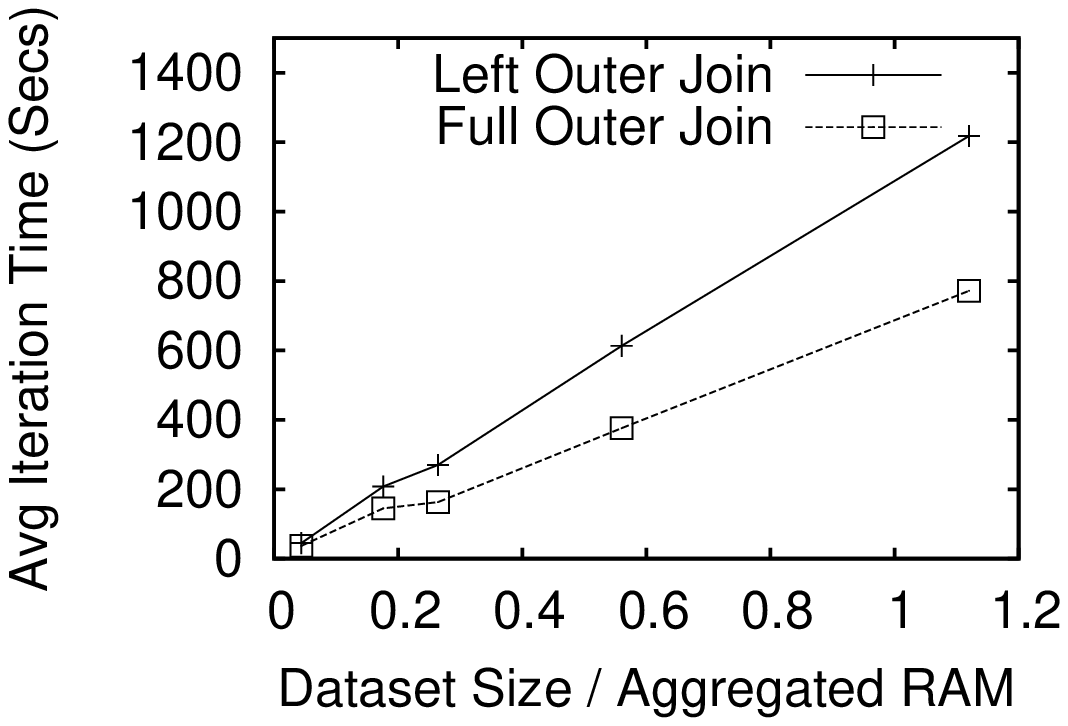}
\end{minipage}
&
\begin{minipage}[t]{0.33\linewidth}
\includegraphics[width=0.90\columnwidth]{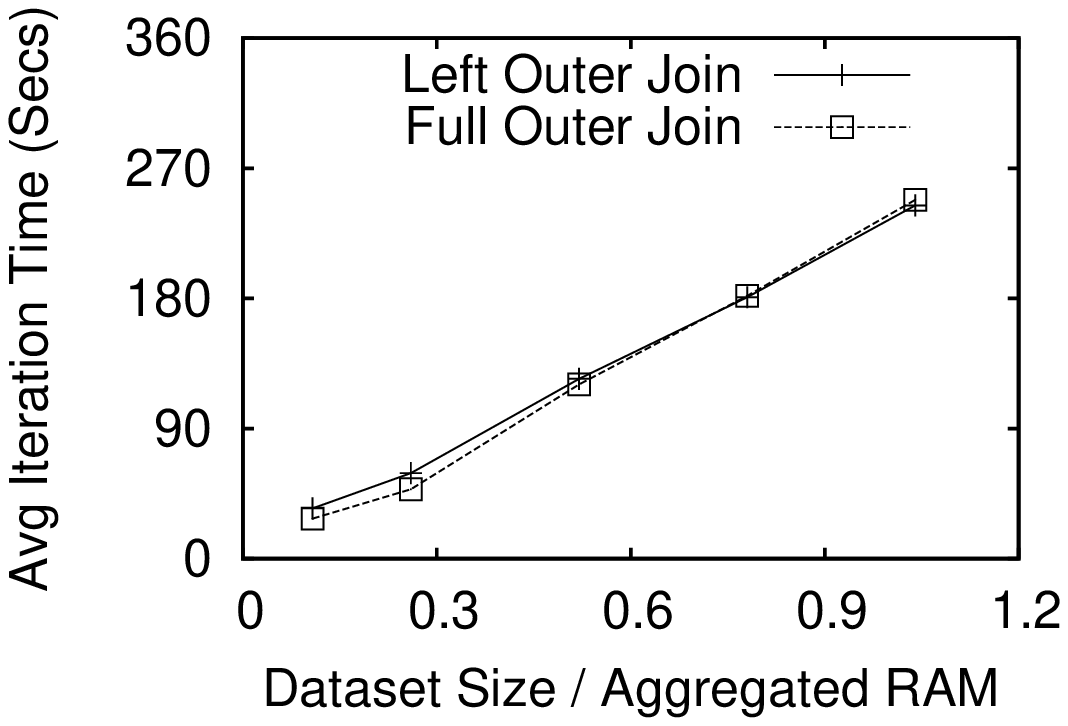}
\end{minipage}
\\
(a) SSSP for BTC datasets  & (b) PageRank for Webmap datasets &(c) CC for BTC datasets\\
\end{tabular}
\vspace{-2ex}
 \caption{Index full outer join vs. index left outer join (run on an 8 machine cluster) for Pregelix.} \label{fig:join}
\vspace{-2ex}
\end{center}
\end{figure*}

\subsection{Plan Flexibility}~\label{flexibility}
\begin{figure}[!t]
\begin{center}
\hspace*{-5ex}
\begin{tabular}{cc}
\begin{minipage}[t]{0.5\linewidth}
\includegraphics[width=\columnwidth]{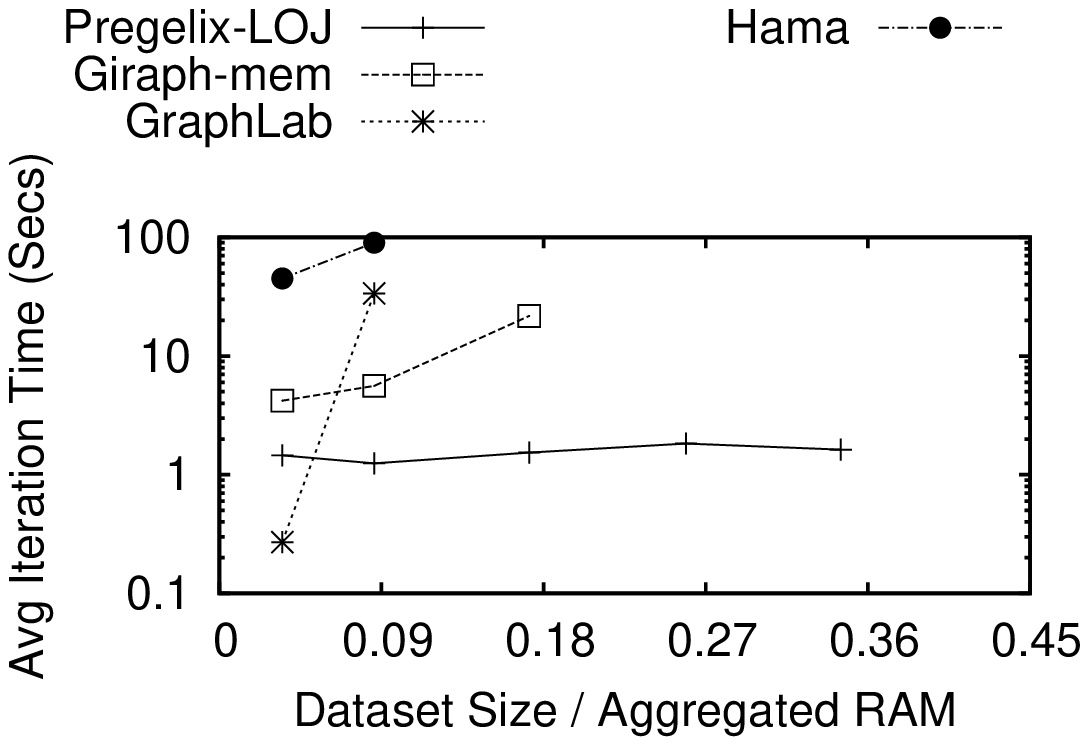}
\end{minipage}
&
\begin{minipage}[t]{0.5\linewidth}
\includegraphics[width=\columnwidth]{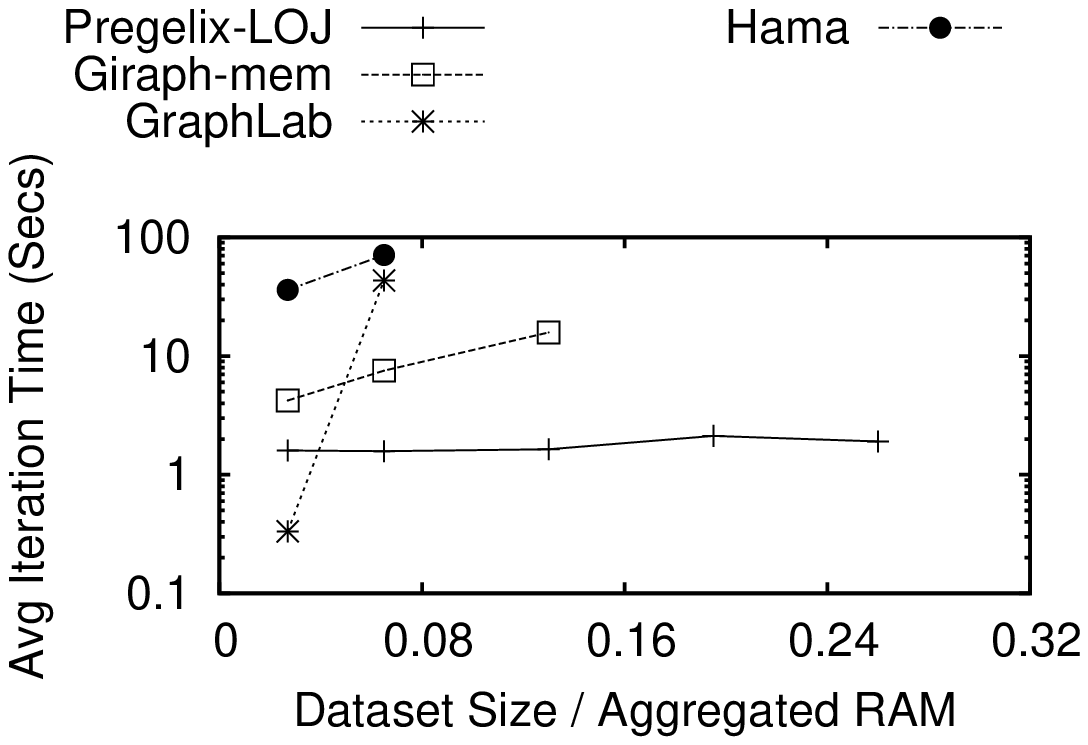}
\end{minipage}\\
(a) 24-Machine Cluster &(b) 32-Machine Cluster \\
\end{tabular}
\vspace{-2ex}
 \caption{Pregelix left outer join plan vs. other systems (SSSP on BTC datasets). GraphX fails to load the smallest dataset.} \label{fig:joinsssp}
\vspace{-2ex}
\end{center}
\end{figure}

In our final experiment, we compare several different physical plan choices in Pregelix to
demonstrate their usefulness. 
We ran the two join plans (described in Section~\ref{joins}) for the three Pregel graph algorithms. 
Figure~\ref{fig:join} shows the results. 
For message-sparse algorithms like SSSP (Figure~\ref{fig:join}(a)),
the left outer join Pregelix plan is much faster than the (default) full outer join plan.
However, for message-intensive algorithms like PageRank (Figure~\ref{fig:join}(b)),
the full outer join plan is the winner. 
This is because although the probe-based left outer join can avoid a sequential index scan,
it needs to search the index from the root node every time;
this is not worthwhile if most data in the leaf nodes will be qualified as join results.
The CC algorithm's execution starts with many messages, but the message volume decreases
significantly in its last few supersteps, and hence the two join plans
result in similar performance (Figure~\ref{fig:join}(c)).
Figure~\ref{fig:joinsssp} revisits the relative performance of the systems by
comparing Pregelix's left outer join plan performance against the other systems.
As shown in the figure, SSSP on Pregelix can be up to 15$\times$ faster than on Giraph 
and up to 35$\times$ faster than on GraphLab for the average per-iteration execution time
when Pregelix is run with its left outer join plan.

In addition to the experiments presented here,  an earlier technical report~\cite{CORR} measured the performance difference introduced by the two different Pregelix data redistribution policies (as described in Section~\ref{groupbys}) for combining messages on a 146-machine cluster in Yahoo! Research.  Figure 9 in the report~\cite{CORR} shows that the m-to-n hash partitioning merging connector can lead to slightly faster executions on small clusters, but
merging input streams at the receiver side needs to selectively wait for data to arrive from specific senders as dictated by the priority queue, and hence it becomes slower on larger clusters.
The tradeoffs seen here and in ~\cite{CORR} for different physical choices are evidence that an optimizer is ultimately essential to identify the best physical plan to use in order to efficiently execute Pregel programs.

\subsection{Software Simplicity}\label{loc}

To highlight the benefit of building a Pregel implementation on top of an existing dataflow runtime,
as opposed to writing one from scratch, we can count lines of code for both the Pregelix and Giraph systems' 
core modules (excluding their test code and comments).
The Giraph-core module, which implements the Giraph infrastructure, contains 32,197 lines of code.
Its counterpart in Pregelix contains just 8,514 lines of code. 

\subsection{Summary}\label{summary}

Our experimental results show that Pregelix can perform comparably to Giraph for
memory-resident message-intensive workloads (like PageRank), can outperform
Giraph by over an order of magnitude for memory-resident message-sparse workloads (like single source shortest
paths), can scale to larger datasets, can sustain multi-user workloads, and also
can outperform GraphLab, GraphX, and Hama by over an order of magnitude on large datasets for various workloads.
In addition to the numbers reported in this paper, an early technical
report~\cite{CORR} gave speedup and scale-up results for the first alpha
release of Pregelix on a 146-machine Yahoo! Research cluster in March 2012;
that study first showed us how well the Pregelix architectural
approach can scale\footnote{Unfortunately, our Yahoo! collaborators have
since left the company, so we no longer have access to a cluster of that scale.}.

\section{Related Work}\label{relatework}
The Pregelix system is related to or built upon
previous works from three areas.

{\bf Parallel data management systems} such as Gamma~\cite{Gamma},
Teradata~\cite{Teradata}, and GRACE~\cite{grace} applied partitioned-parallel
processing to SQL query processing over two decades ago.  The introduction of
Google's MapReduce system~\cite{MapReduce}, based on similar principles, led to
the recent flurry of work in MapReduce-based data-intensive computing.  Systems
like Dryad~\cite{Dryad}, Hyracks~\cite{Hyracks}, and Nephele~\cite{Nephele}
have successfully made the case for supporting a richer set of data operators
beyond map and reduce as well as a richer set of data communication patterns.
REX~\cite{REX} integrated user-defined delta functions into SQL to support
arbitrary recursions and built stratified parallel evaluations for recursions.
The Stratosphere project also proposed an incremental iteration
abstraction~\cite{WorkingSet} using working set management and integrated it
with parallel dataflows and job placement.  The lessons and experiences from
all of these systems provided a solid foundation for the Pregelix system.

{\bf Big Graph processing platforms} such as Pregel~\cite{Pregel},
Giraph~\cite{Giraph}, and Hama~\cite{Hama}
have been built to provide vertex-oriented message-passing-based
programming abstractions for distributed graph algorithms
to run on shared-nothing clusters. Sedge~\cite{Sedge}
proposed an efficient advanced partitioning scheme to minimize inter-machine
communications for Pregel computations.  Surfer~\cite{Bandwidth} 
is a Pregel-like prototype using advanced bandwidth-aware graph partitioning to minimize the network traffic in
processing graphs. 
In seeming contradiction to these favorable results on the efficacy of smart partitioning, literature~\cite{LFGraph} found that basic hash partitioning works better
because of the resulting balanced load and the low partitioning overhead.
We seem to be seeing the same with respect to GraphLab, i.e., the pain is not being repaid in performance gain.
GPS~\cite{GPS} optimizes Pregel computations by
dynamically repartitioning vertices based on message patterns
and by splitting high-degree vertices across all
worker machines.  Giraph++~\cite{Giraph++} enhanced Giraph with a richer set of APIs
for user-defined partitioning functions so that communication 
within a single partition can be bypassed. 
GraphX~\cite{GraphX} provides a programming abstraction called Resident Distributed Graphs (RDGs) 
to simplify graph loading, construction, transformation, and computations,  on top of
which Pregel can be easily implemented.
Different from Pregel, GraphLab~\cite{GraphLab} provides a vertex-update-based programming abstraction
and supports an asynchronous model to increase the level of pipelined parallelism.
Trinity~\cite{Trinity} is a distributed graph processing engine built on top of a distributed in-memory key-value store
to support both online and offline graph processing; it optimizes message-passing in 
vertex-centric computations for the case where a
vertex sends messages to a fixed set of
vertices.
Our work on Pregelix is largely orthogonal to these systems and their contributions
because it looks at Pregel at a lower architectural level, aiming
at better out-of-core support, plan flexibility, and software simplicity.

{\bf Iterative extensions to MapReduce} like HaLoop~\cite{haloop} and
PrIter~\cite{PrIter} were the first to extend MapReduce with looping constructs.  HaLoop hardcodes a sticky scheduling policy (similar to 
the one adopted here in Pregelix and to the one in Stratosphere~\cite{WorkingSet}) 
into the Hadoop task scheduler so as to introduce a caching ability for iterative analytics.  PrIter uses a
key-value storage layer to manage its intermediate MapReduce state, and it also
exposes user-defined policies that can prioritize certain data to promote fast
algorithmic convergence.  However, those extensions still constrain
computations to the MapReduce model, while Pregelix explores more
flexible scheduling mechanisms, storage options, 
operators, and several forms of data redistribution (allowed by Hyracks) to 
optimize a given Pregel algorithm's computation time.

%
%


\section{Conclusions}\label{conclusions}
This paper has presented the design, implementation, early use cases, and evaluation 
of Pregelix, a new dataflow-based Pregel-like system
built on top of the Hyracks parallel dataflow engine.
Pregelix combines the Pregel API from the systems world
with data-parallel query evaluation techniques from the database world
in support of Big Graph Analytics.
This combination leads to effective and transparent out-of-core support, scalability, and throughput, as well as increased software simplicity and physical flexibility.
To the best of our knowledge, Pregelix is the only open source Pregel-like
system that scales to out-of-core workloads efficiently, can sustain multi-user workloads, and allows runtime flexibility.
This sort of architecture and methodology could be adopted by parallel data warehouse
vendors (such as Teradata~\cite{Teradata}, Pivotal~\cite{Pivotal}, or Vertica~\cite{Vertica}) to build Big Graph
processing infrastructures on top of their existing query execution engines. 
Last but not least, we have made several stable releases of the Pregelix
system (http://pregelix.ics.uci.edu) in open source 
form since the year 2012 for use by the Big Data research community,
and we invite others to download and try the system.
As future work, we plan to automate physical
plan selection via a cost-based optimizer (similar to literature~\cite{MROptimizer}) and we plan to integrate
Pregelix with AsterixDB~\cite{AsterixDB} to support richer forms of
Big Graph Analytics.

\subsection*{Acknowledgements}\label{acknowledgement}
{
The Pregelix project has been supported by a
UC Discovery grant, NSF IIS award 0910989 and 1302698, and NSF CNS awards 1305430, 1351047, and 1059436.
Yingyi Bu is supported in part by a Google Fellowship award.
The AsterixDB project has enjoyed industrial support from Amazon, eBay,
Facebook, Google, HTC, Microsoft, Oracle Labs, and Yahoo!.
We also thank the following people who tried Pregelix at the early stage, reported
bugs, hardened the system, and gave us feedback: Jacob Biesinger, Da Yan,  Anbang Xu,
Nan Zhang, Vishal Patel, Joe Simons, Nicholas Ceglia, Khanh Nguyen, Hongzhi Wang, James Cheng, Chen Li, Xiaohui Xie, 
and Harry Guoqing Xu.
Finally, we thank Raghu Ramakrishnan for the early discussion of this work as well
as the sponsorship for our access to the Yahoo! cluster for scale-testing
early versions of the system.
}

\bibliographystyle{abbrv}
{
\vspace{-2ex}
\scriptsize
\bibliography{reference}
}


\end{document}